\documentclass[12pt]{article}

\usepackage[english]{babel}
\usepackage{amsmath}		 		
\usepackage{mathtools}
\usepackage{url}					   
\usepackage[utf8x]{inputenc}	 
\usepackage{natbib}				     
\usepackage{parskip}				 
\usepackage{fancyhdr}				
\usepackage{vmargin}				
\usepackage{fixltx2e}				  

\usepackage{array}						
\usepackage{caption}  				
\usepackage{chngcntr}				
\usepackage{hhline}					
\usepackage{multirow}				
\usepackage{booktabs,caption}	
\usepackage[flushleft]{threeparttable}	 

\counterwithin{table}{section}			

\newcolumntype{L}[1]{>{\raggedright\let\newline\\\arraybackslash\hspace{0pt}}m{#1}}
\newcolumntype{C}[1]{>{\centering\let\newline\\\arraybackslash\hspace{0pt}}m{#1}}
\newcolumntype{R}[1]{>{\raggedleft\let\newline\\\arraybackslash\hspace{0pt}}m{#1}}

\usepackage{graphicx}				
\usepackage{subfig}					
\usepackage{float}						
\graphicspath{{images/}}				

\counterwithin{figure}{section}	

\counterwithin{equation}{section}		

\setmarginsrb{2.5 cm}{2 cm}{2.5 cm}{2 cm}{1 cm}{1.5 cm}{1 cm}{1.5 cm}

\usepackage{xcolor}
\usepackage{listings}             

\definecolor{codegreen}{rgb}{0,0.6,0}
\definecolor{codegray}{rgb}{0.5,0.5,0.5}
\definecolor{codepurple}{rgb}{0.58,0,0.82}
\definecolor{backcolour}{rgb}{0.95,0.95,0.92}

\lstdefinestyle{customf}
{
	backgroundcolor=\color{backcolour},
	commentstyle=\color{codegreen},
	keywordstyle=\color{blue},
	identifierstyle=\color{black},
	breakatwhitespace=false,
	basicstyle=\footnotesize\ttfamily,
	breaklines=true,
	captionpos=b,
	keepspaces=true,
	showspaces=false,
	showstringspaces=false,
	showtabs=false,
	tabsize=2
}

\usepackage[skip=0pt]{caption}

\usepackage{tcolorbox}

\tcbuselibrary{listings}
\newtcblisting{ccode}
{
	colframe = black,
	boxrule=0.5pt,
	listing only,
	listing options={language=C++}
}

\newtcblisting{shcode}
{
	colframe = black,
	boxrule=0.5pt,
	listing only,
	listing options={language=bash}
}

\title{Grid2Grid : HOS Wrapper Program for CFD solvers }

\begin{document}


\fancyhf{}
\cfoot{\thepage}

\makeatletter
\let\thetitle\@title
\let\theauthor\@author
\let\thedate\@date
\makeatother


\begin{titlepage}
	\centering
    \vspace*{0.5 cm}    

	\rule{\linewidth}{0.2 mm} \\[0.8 cm]
	{ \huge \bfseries \thetitle}\\[0.3 cm]
	\rule{\linewidth}{0.2 mm} \\[0.8 cm]
	\textup{\large YoungMyung Choi, Maite Gouin, Guillaume Ducrozet, }\\[0.5 cm]		
	\textup{\large Benjamin Bouscasse and Pierre Ferrant}\\[1.0 cm]		
	
	\large{17 November, 2017} \\[0.5cm]	
	\large{Version 1.0} \\[4.0cm]

	\includegraphics[scale=0.4]{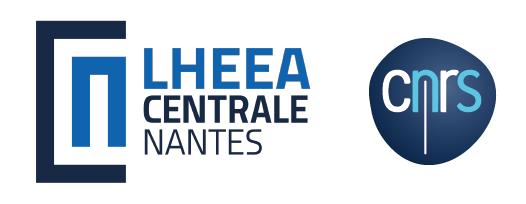}  \\[1.5 cm]
	
	 \textsc{\LARGE Ecole Centrale de Nantes}\\[1.0 cm]	
	 
\end{titlepage}

	\tableofcontents	
\relax 
\pagebreak
\section{Introduction}

Wave generation solvers using Higher Order Spectral Method (HOS) have been validated and developed for several years (\cite{ducrozet2016}, \cite{ducrozet2007}, \cite{bonnefoy2011} and \cite{ducrozet2012}). HOS solves nonlinear wave propagation in open-sea (HOS-Ocean) and also in numerical wave tank (HOS-NWT) with low computation time comparing with other nonlinear wave solvers because the theory is based on pseudo-spectral method (\cite{HOSOcean} and \cite{HOSNWT}). Those HOS wave solvers are released as open-source codes, which anyone can develop, use and distribute under the terms of GNU General Public Licence (GPLv3).

Nonlinear irregular wave generation in computational fluids dynamic (CFD) solvers becomes important recently in naval fields to better estimate the loads on offshore structure. The conventional linear superposition methods imply a long computational time for wave generation and the simulation is made almost impossible without having enough computational power. And if the method is based on linear wave theory, there is also question on occurence of nonlinear phenomenon and the interaction between waves as the simulation goes. Therefore dedicated nonlinear wave solvers with high computational speed are needed.

\texttt{Grid2Grid} is developed as a wrapper program of HOS to generate wave fields from the results of HOS computation. \texttt{Grid2Grid} reconstructs wave fields of HOS with inverse fast Fourier transforms (FFTs) and uses a quick spline module, the nonlinear wave fields can be fastly reconstructed for arbitrary simulation time and space. The nonlinear wave simulation is then possible for a particular position and time where specific non linear phenomenon occur.

\texttt{Grid2Grid} compiles a shared library (\texttt{libGrid2Grid.so}) which can be used for communication with other progamming language using the \texttt{ISO\_C\_BINDING} rule. It compiles also a post processing program of HOS.

\pagebreak
\section{Program Structure}

The program structure of \texttt{Grid2Grid} is depicted in Fig. \ref{fig:grid2GridStructure}. The program is composed of several modules (\texttt{.f90}) which are denoted as \texttt{mod}. HOS wave fields are generated by  \texttt{modSurf2Vol} from the HOS result file (\texttt{modes\_HOS\_SWENSE.dat}). Because the results files only contains modes information, the volumic wave fields is reconstructed on HOS grid by inverse FFTs (denoted as $H_2$ Operator). The volumic wave fields in \texttt{Grid2Grid} are called as snapshot of wave fields because it is re-constructed at certain simulation time and grid of HOS. \texttt{modVol2Vol} constructs the interpolation data structure from several snapshot of wave fields by using multidimensional spline module (\cite{BsplineWilliams}). It can directly communicate with other Fortran language. To communicate with other languages, \texttt{modGrid2Grid} can also be used as it receives input data and return wave fields data by using \texttt{ISO\_C\_BINDING}. \texttt{modPostGrid2Grid} is a post processing module of HOS generating wave fields in VTK format and also wave elevation time series.
\newline

{
    \begin{figure} [H]
    \centering
    \includegraphics[scale=0.62]{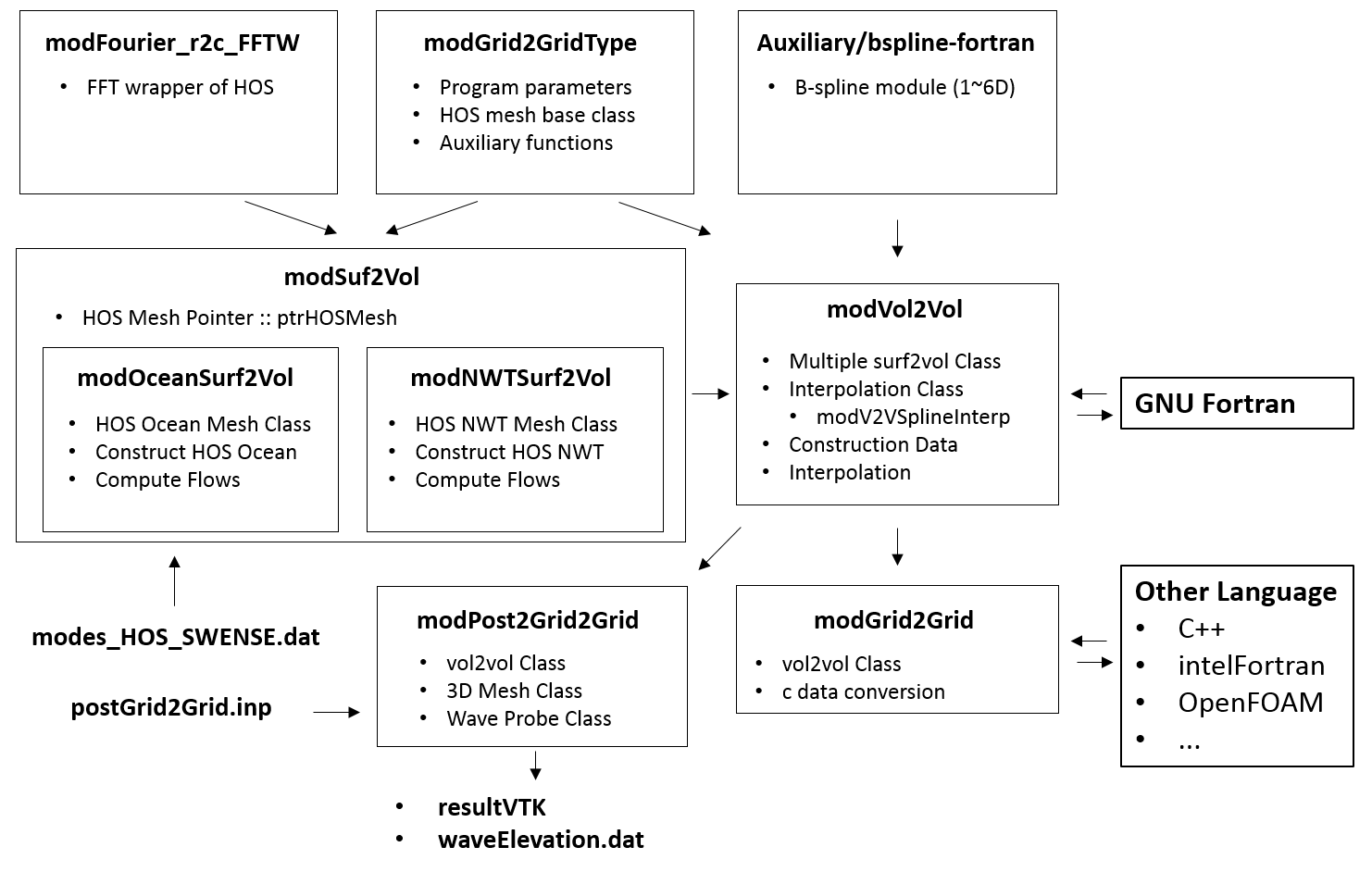}
    \caption{Program Structure of \texttt{Grid2Grid}}
    \label{fig:grid2GridStructure}
    \end{figure}
}

\pagebreak

The main feature of each module is following.
\newline

{
\hspace{0.5 cm} -- \texttt{modSurf2Vol} : Wrapper class of HOS-Ocean and HOS-NWT wave fields

\hspace{0.5 cm} -- \texttt{modVol2Vol} : Interpolation of snapshot of wave fields (multiple Surf2Vol Class)

\hspace{0.5 cm} -- \texttt{modGrid2Grid} : Communication with other language with \texttt{ISO\_C\_BINDING}

\hspace{0.5 cm} -- \texttt{modPostGrid2Grid} : Post processing of HOS (wave fields and wave elevation)

\hspace{0.5 cm} -- \texttt{modOceanSurf2Vol} : Re-construction HOS-Ocean wave fields

\hspace{0.5 cm} -- \texttt{modNWTSurf2Vol} : Re-construction HOS-NWT wave fields

\hspace{0.5 cm} -- \texttt{modFourier\_r2c\_FFTW} : Wrapper module of \texttt{FFTW} open library

\hspace{0.5 cm} -- \texttt{modGrid2GridType} : Grid2Grid global variables and auxiliary functions

\hspace{0.5 cm} -- \texttt{bspline-fortran} : Multidimensional b-spline module written in Fortran.
}

\pagebreak	
\subsection{typSurf2Vol}

\subsubsection{Description}

\texttt{Surf2Vol} is wrapper class to access HOS Ocean and NWT class. Functionality and data of both HOS classes are similar but the mathematical formulation and meshes are a little bit different. Therefore the flow information from each HOS grid is only taken by using HOS mesh base class and transfer it to \texttt{Vol2Vol} Class. It is reminded that the direct access on the HOS wave fields is only possible by HOS mesh pointer to prevent misuse of HOS wave data. 

\texttt{Surf2Vol} class structure is depicted in Fig. \ref{fig:surf2VolStructure}. \texttt{Surf2Vol} class contains \texttt{HOSOceanSurf2Vol} and \texttt{HOSNWTSurf2Vol} classes. When the functionality is called, it calls the functionality of sub-class distiguished by HOS type. When the subroutine \texttt{correct} is called, the sub-class reads HOS modes from HOS result file and conducts $H_2$ operation to re-construct wave fields. Reconstructed wave field is saved on derived HOS grid class. And the HOS grid pointer at \texttt{Surf2Vol} points to derived HOS grid class to access wave fields. 

\vspace{0.5cm}
{
	\begin{figure} [H]
		\centering
		\includegraphics[scale=0.78]{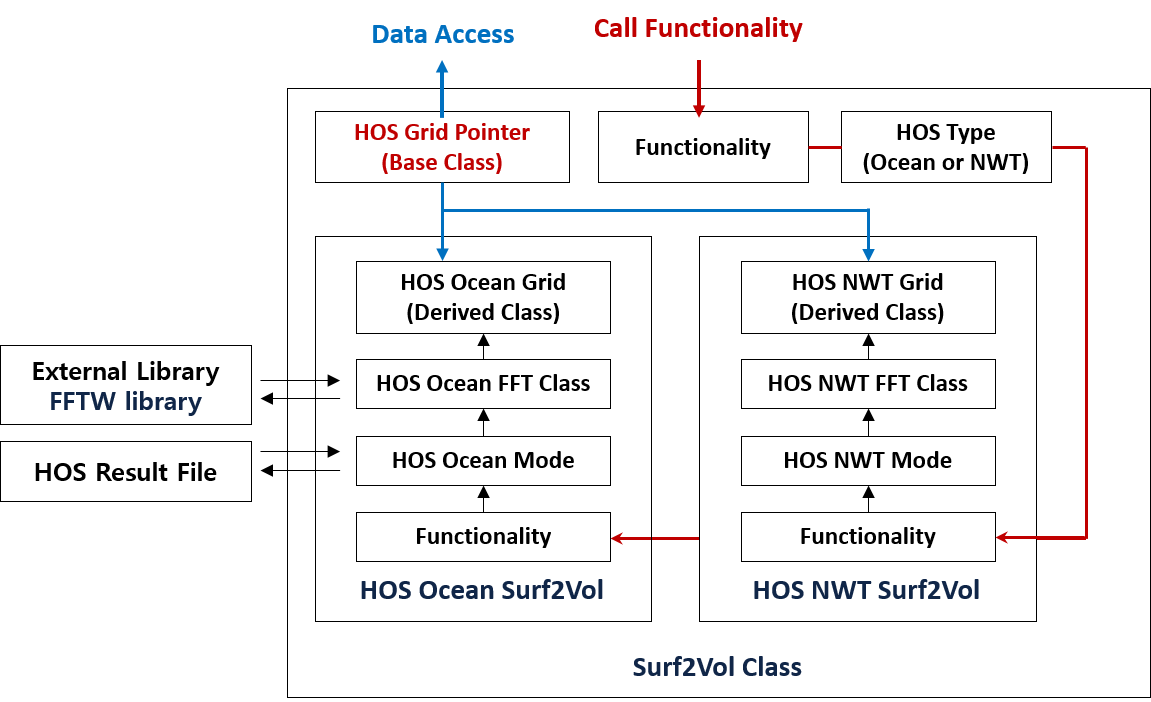}
		\vspace{0.5cm}
		\caption{\texttt{Surf2Vol} class structure}
		\label{fig:surf2VolStructure}
	\end{figure}
}

\pagebreak

\texttt{hosNWTSurf2Vol} and \texttt{hosOceanSurf2Vol} classes reconstruct HOS wave fields from HOS result file by using the $H_2$ operator. Wave reconstruction and HOS wave theory is well explained in \cite{ducrozet2007} and \cite{ducrozet2012}. Each class constructs wave fields on derived HOS mesh class which can be pointed by HOS mesh base class. 

HOS wave theory is based on the superposition of base modes which satisfy Laplace equation, sea bottom and free surface boundary condition basically, the behavior of the mode function has an exponential profile. This exponential profile could magnify local modes which has high wave number above $z=0$. For high steepness waves, this property gives unnatural values closed to free surface, criterion to cut local mode is introduced as \eqref{eq:fnzCriterion}. 

{
	\begin{equation}
	\label{eq:fnzCriterion}
	f_{mn}(z) = \begin{cases}
	\dfrac{\cosh k_{mn}(z+H)}{\cosh k_{mn}H}  &  \text{if} \quad f_{mn}(z) < C_{f(z)} \\
	C_{f(z)}  &  \text{if} \quad f_{mn}(z) \geq C_{f(z)}
	\end{cases}
	\end{equation}
	\newline
	\centering
	where
	\begin{equation*}
	k_{mn}	\text{ : HOS mode wave number \quad  $z$ : HOS z-coordinates \quad $H$ :  water depth}
	\end{equation*} 
	\begin{equation*}
	C_{f(z)}  \text{ : $f_{mn}(z)$ function criterion value (default = 10) }
	\end{equation*}
}

The parameter $C_{f(z)}$ is set to the default value of  10 in \texttt{Grid2Grid}. If the local wave velocities generated by \texttt{Grid2Grid} are not sufficient, the parameter can be changed. The parameter is defined in \texttt{modGrid2GridType.f90} as a \texttt{FNZ\_VALUE}.

HOS result file only contains mode amplitudes computed at each HOS simulation time, the volumic grid should then be reconstructed from those modes. HOS does not need vertical grid information, so z-directional grid information should be given when HOS \texttt{surf2vol} is initialised. Constructed HOS grid is used for interpolation grid. 

To constuct the HOS grid, following information should be given.

\hspace{0.5 cm} $\circ$ \texttt{zMin}, \texttt{zMax}

\hspace{0.5 cm} $\circ$ \texttt{nZMin}, \texttt{nZMax}

\hspace{0.5 cm} $\circ$ \texttt{nZMinRatio}, \texttt{nZMaxRatio} [Optional]

where \texttt{zMin} and \texttt{zMax} is used to set $z$-directional domain and it should have negative and positive value. \texttt{nZMin} and \texttt{nZMax} are the number of $z$-directional grid. It is recommended to have at least 50-100 points for the interpolation. If a sufficient number of grid points is not given, the interpolation scheme could give strange values due to exponetial profile of $f_{mn}(z)$. \texttt{nZMinRatio} and \texttt{nZMaxRatio} are the ratio of maximum and minumum height of grid (${\Delta z_{max}}/{\Delta z_{min}} $). Minimum grid is located at $z=0$. Those are optional values set to be 3 as default.  The grid can be visualized by using ParaView. The VTK file is located at VTK/Grid2Grid/surf2volMesh.vtk.

\pagebreak
\subsubsection{Class (Type)}

\textbf{Class} : \texttt{Surf2Vol}

\hspace{0.5 cm} -- Data :

\hspace{1.0 cm} $\circ$ \texttt{hosNWTSurf2Vol\_} : HOS-NWT Surf2Vol Class

\hspace{1.0 cm} $\circ$ \texttt{hosOceanSurf2Vol\_} : HOS-Ocean Surf2Vol Class

\hspace{1.0 cm} $\circ$ \texttt{ptrHOSMesh\_} : HOS Mesh Pointer\\

\hspace{0.5 cm} -- Functionality :

\hspace{1.0 cm} $\circ$ \texttt{initialize} : Initialise HOS Surf2Vol class with HOS type, result file path, ...

\hspace{1.0 cm} $\circ$ \texttt{correct} : Update HOS wave fields

\hspace{1.0 cm} $\circ$ \texttt{destroy} : Class desroyer\\    

\vspace{0.1cm}

\textbf{Class} : \texttt{HOSOceanSurf2Vol} and \texttt{HOSNWTSurf2Vol}

\hspace{0.5 cm} -- Data :

\hspace{1.0 cm} $\circ$ \texttt{HOSfile} : HOS result file (\texttt{modes\_HOS\_SWENSE.dat})

\hspace{1.0 cm} $\circ$ \texttt{HOSmode} : HOS Ocean or NWT modes

\hspace{1.0 cm} $\circ$ \texttt{HOSmesh} : HOS Ocean or NWT mesh (Two are different)

\hspace{1.0 cm} $\circ$ \texttt{HOSfftw} : HOS FFT class for $H_2$ operator \\

\hspace{0.5 cm} -- Public Functionality :

\hspace{1.0 cm} $\circ$ \texttt{initialize} : Initialise HOS Ocean or NWT Surf2Vol with result file path, ...

\hspace{1.0 cm} $\circ$ \texttt{correct} : Update HOS Ocean or NWT wave fields

\hspace{1.0 cm} $\circ$ \texttt{destroy} : Class desroyer\\

\hspace{0.5 cm} -- Private Functionality :

\hspace{1.0 cm} $\circ$ \texttt{init\_read\_mod} : Read HOS number of modes and allocate dynamic array

\hspace{1.0 cm} $\circ$ \texttt{read\_mod} : Read HOS modes for given HOS time index

\hspace{1.0 cm} $\circ$ \texttt{buildGlobalMesh} : Build HOS mesh with given HOS construction parameter

\hspace{1.0 cm} $\circ$ \texttt{reconstuctionFFTs} : Reconstruct wave fields by inverse $H_2$ operator \\

\subsubsection{How to use}

\hspace{0.5 cm} -- Initialise \texttt{Surf2Vol}

\begin{lstlisting}[language={[95]Fortran}]

Call hosS2V%initialize(hosType, filePath, zMin, zMax, nZmin, nZmax, zMinRatio, zMaxRatio)

!	hosS2V				: Surf2vol Class (Type)
!	hosType				: HOS Type (Ocean or NWT)
!	filePath			: HOS result file path (modes_HOS_SWENS.dat)
!	zMin, zMax 		: HOS grid z-minimum and z-maximul (vertical domain)
!	nZmin, nZmax 	: Number of z-directional Grid	
!
!	zMinRatio, zMaxRatio (Optional)
!	: Ratio of maximum and minimum height of grid (default=3)
\end{lstlisting}		

\vspace{0.5cm}	

\hspace{0.5 cm} -- Correct \texttt{Surf2Vol}

\begin{lstlisting}[language={[95]Fortran}]

Call hosS2V%correct(hosIndex)

!	hosS2V				: Surf2vol Class (Type)
!	hosIndex			: HOS time index
\end{lstlisting}		

\vspace{0.5cm}	

\hspace{0.5 cm} -- Data access on wave field of \texttt{Surf2Vol}

\hspace{1.0 cm} $\circ$ Wave elevation 

\begin{lstlisting}[language={[95]Fortran}]

eta = hosS2V%ptrHOSMesh_%eta(ix, iy)

!	hosS2V				: Surf2vol class (Type)
!	ix, iy				: HOS grid index (x, y)
!	eta						: Wave elevation
\end{lstlisting}	

\hspace{1.0 cm} $\circ$ Wave velocity

\begin{lstlisting}[language=bash]

u = hosS2V%ptrHOSMesh_%eta(ix, iy, iz)
v = hosS2V%ptrHOSMesh_%eta(ix, iy, iz)
w = hosS2V%ptrHOSMesh_%eta(ix, iy, iz)

!	hosS2V				: Surf2vol class (Type)
!	ix, iy, iz		: HOS grid index (x, y, z)
!	u, v, w				: Wave velocity (x, y, z)
\end{lstlisting}	

\pagebreak

\hspace{1.0 cm} $\circ$ Dynamic pressure ($p_d = p - \rho g z$)

\begin{lstlisting}[language={[95]Fortran}]

pd = hosS2V%ptrHOSMesh_%pd(ix, iy, iz)

!	hosS2V				: Surf2Vol class (Type)
!	ix, iy, iz		: HOS grid index (x, y, z)
!	pd						: Dynamic pressure
\end{lstlisting}	

\vspace{0.5cm}	

\hspace{0.5 cm} -- Destruct of \texttt{Surf2Vol}

\begin{lstlisting}[language={[95]Fortran}]

Call hosS2V%destroy()

!	hosS2V				: Surf2Vol class (Type)
\end{lstlisting}

	\pagebreak
	\subsection{typVol2Vol}

	\subsubsection{Description}

	\texttt{Vol2Vol} is an interpolation class used to give interpolated wave information data from the reconstructed HOS wave field. It holds several \texttt{Surf2Vol} classes and interpolation class. The \texttt{Vol2Vol} class structure is described in Fig \ref{fig:vol2volStructure}. 

	HOS result file holds modes amplitudes time series for the whole HOS simulation time. If we construct HOS wave fields and interpolation data structure for the whole HOS simulation time, not only the computation time is long but also a huge memory is demanded. \texttt{Grid2Grid} aims for construction of demanding wave fields for relatively short period and domain, it is not necessary to construct the whole HOS wave fields and interpolation data structure. Therefore revolving algorithm is applied just to update HOS wave fields adjucent to simulation time and constuct small interpolation data structure for efficient computation time and memory.  

	When \texttt{initialize} of \texttt{Vol2Vol} is called, it allocates \texttt{Surf2Vol} array, interpolation data and array related to revolving algorithm. And it initialises \texttt{Surf2Vol} classes and interpolation class and call its subroutine \texttt{correct} at $t=0$.

	{
		\begin{figure} [H]
			\centering
			\includegraphics[scale=0.72]{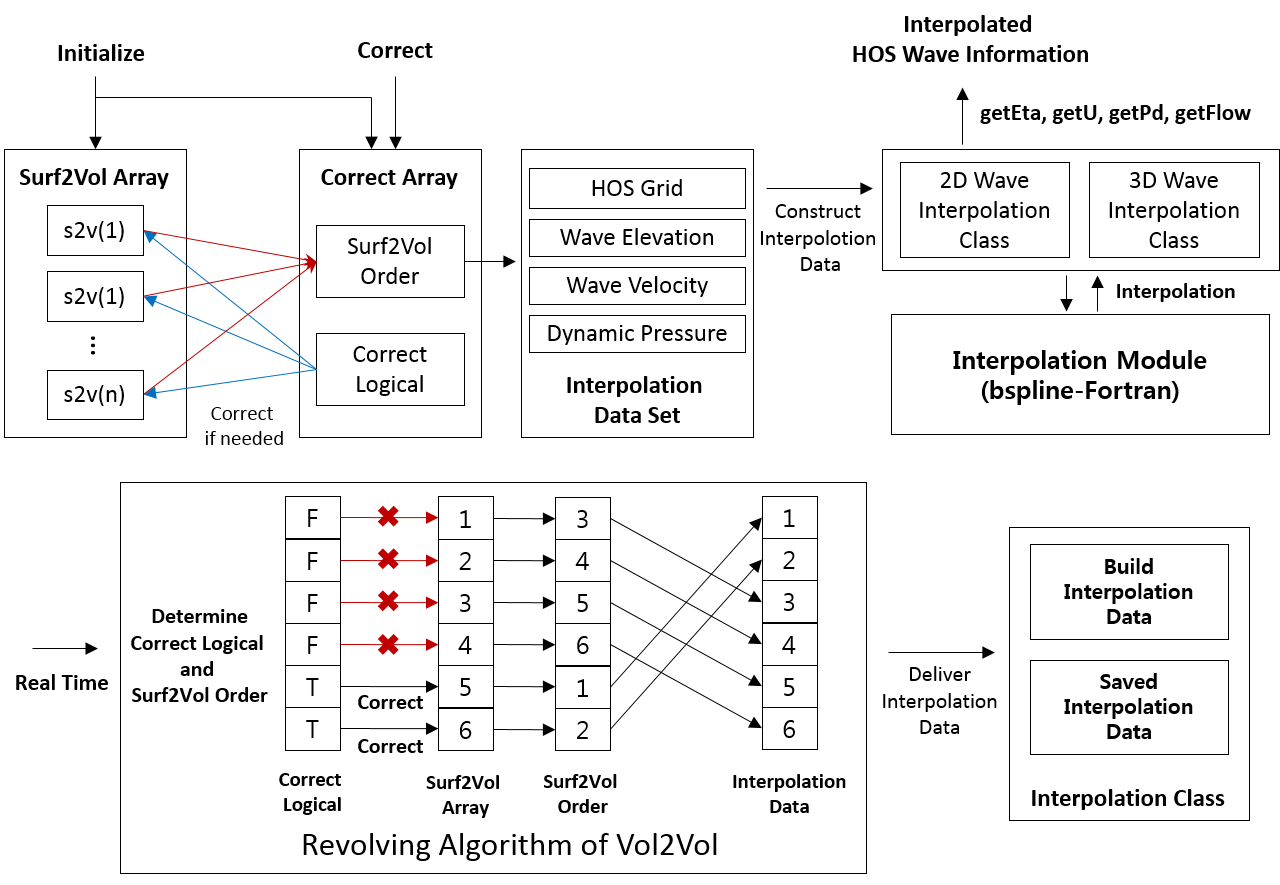}
			\vspace{0.2cm}
			\caption{\texttt{Vol2Vol} class structure}
			\label{fig:vol2volStructure}
		\end{figure}
	}

	\pagebreak

	The subroutine \texttt{correct} with input $t$ first determine HOS \texttt{Surf2Vol} correction index and \texttt{Surf2Vol} order based on input $t$ and previous HOS time index. By using correction index and order, only necessary HOS \texttt{Surf2Vol} is updated and re-ordered to constructed for interpolation data. After interpolation data is constructed, it is delivered to the interpolation class. The interpolation class communicates with the \texttt{bspline-Fortran} module and constructs the interpolation data structure. 

	The subroutine \texttt{getEta}, \texttt{getU}, \texttt{getPd} and \texttt{getFlow} return interpolated values for given space and time from the constructed interpolation data structure.

	\subsubsection{Class (Type)}
	
	\textbf{Class} : \texttt{Vol2Vol}
	
	\hspace{0.5 cm} -- Data :
	
	\hspace{1.0 cm} $\circ$ \texttt{nInterp\_} : Interpolation order (2 : Linear, 3 : Quadratic, 4 : Cubic, ... )
	
	\hspace{1.0 cm} $\circ$ \texttt{nSaveT\_} : Number of Surf2Vol wave fields  (\texttt{nSaveT\_}  $>$ \texttt{nInterp\_})
	
	\hspace{1.0 cm} $\circ$ \texttt{HOSs2v\_(:)} : Array of \texttt{Surf2Vol} class
	
	\hspace{1.0 cm} $\circ$ \texttt{itp2D\_} : Interpolation class for 2D waves
	
	\hspace{1.0 cm} $\circ$ \texttt{itp3D\_} : Interpolation class for 3D waves
	
	\vspace{0.5cm}
	
	\hspace{0.5 cm} -- Functionality :
	
	\hspace{1.0 cm} $\circ$ \texttt{initialize} : Initialise HOS Vol2Vol class with HOS type, result file path, ...
	
	\hspace{1.0 cm} $\circ$ \texttt{correct} : Update HOS wave fields with real-time
	
	\hspace{1.0 cm} $\circ$ \texttt{getEta} : Get interpolated wave elevation 
	
	\hspace{1.0 cm} $\circ$ \texttt{getU} : Get interpolated wave velocity 
	
	\hspace{1.0 cm} $\circ$ \texttt{getPd} : Get interpolated dynamic pressure
	
	\hspace{1.0 cm} $\circ$ \texttt{getFlow} : Get flow information
	
	\hspace{1.0 cm} $\circ$ \texttt{destroy} : Class destroyer\\    
	
	\pagebreak
	
	\subsubsection{How to use}
	
	\hspace{0.5 cm} -- Initialise \texttt{Vol2Vol}
	
	\begin{lstlisting}[language={[95]Fortran}]
	
	Call hosV2V%initialize(hosType, filePath, zMin, zMax, nZmin, nZmax, zMinRatio, zMaxRatio, iflag)
	
	!	hosV2V				: Vol2vol Class (Type)
	!	hosType				: HOS Type (Ocean or NWT)
	!	filePath			: HOS result file path (modes_HOS_SWENS.dat)
	!	zMin, zMax 		: HOS grid z-minimum and z-maximul (vertical domain)
	!	nZmin, nZmax 	: Number of z-directional Grid	
	!
	!	zMinRatio, zMaxRatio  (Optional)
	!	: Ratio of maximum and minimum height of grid (default=3)
	!
	!	iflag : Wriging option (iflag = 1, write Grid2Grid information)
	\end{lstlisting}
	
	\hspace{0.5 cm} -- Correct \texttt{Vol2Vol}
	
	\begin{lstlisting}[language={[95]Fortran}]
	
	Call hosV2V%correct(simulTime)
	
	!	hosV2V				: Vol2vol Class (Type)
	!	simulTime			: Simulation time (real time value)
	\end{lstlisting}	
	
	\hspace{0.5 cm} -- Get wave elevation from \texttt{Vol2Vol}
	
	\begin{lstlisting}[language={[95]Fortran}]
	
	eta = hosV2V%getEta(x, y, simulTime, iflag)
	
	!	hosV2V				: Vol2vol Class (Type)
	!	x, y					: x and y position 
	!	simulTime 		: Simulation time (real time value)
	!
	!	eta						: Wave elevation
	!
	!	iflag (Optional)	
	!	: if iflag = 1, nondimensional x and y can be given (default = 0). 	
	\end{lstlisting}		
	
	\pagebreak	
	
	\hspace{0.5 cm} -- Get wave velocity from \texttt{Vol2Vol}
	
	\begin{lstlisting}[language={[95]Fortran}]
	
	Call hosV2V%getU(x, y, z, simulTime, u, v, w, iflag)
	
	!	hosV2V				: Vol2vol Class (Type)
	!	x, y, z				: x, y and z position 
	!	simulTime 		: Simulation time (real time value)
	!
	!	u, v, z				: Wave velocity ( x, y, z )
	!
	!	iflag (Optional)	
	!	: if iflag = 1, nondimensional x and y can be given (default = 0). 	
	\end{lstlisting}	
	
	\hspace{0.5 cm} -- Get dynamic pressure from \texttt{Vol2Vol}
	
	\begin{lstlisting}[language={[95]Fortran}]
	
	pd = hosV2V%getPd(x, y, z, simulTime, iflag)
	
	!	hosV2V				: Vol2vol Class (Type)
	!	x, y, z				: x, y and z position 
	!	simulTime 		: Simulation time (real time value)
	!
	!	pd						: Dynamic velocity (pd = p - rho * g * z)
	!
	!	iflag (Optional)	
	!	: if iflag = 1, nondimensional x and y can be given (default = 0). 	
	\end{lstlisting}	

	\hspace{0.5 cm} -- Get flow information from \texttt{Vol2Vol}
	
	\begin{lstlisting}[language={[95]Fortran}]
	
	Call hosV2V%getFlow(x, y, z, simulTime, eta, u, v, w, pd, iflag)
	
	!	hosV2V				: Vol2vol Class (Type)
	!	x, y, z				: x, y and z position 
	!	simulTime 		: Simulation time (real time value)
	!
	!	eta						: Wave elevation
	!	u, v, z				: Wave velocity ( x, y, z )
	!	pd						: Dynamic velocity (pd = p - rho * g * z)
	!
	!	iflag (Optional)	
	!	: if iflag = 1, nondimensional x and y can be given (default = 0). 	
	\end{lstlisting}	
	
	\hspace{0.5 cm} -- Destroy \texttt{Vol2Vol}
	
	\begin{lstlisting}[language={[95]Fortran}]

	Call hosV2V%destroy()	
	\end{lstlisting}	

\pagebreak
\subsection{modGrid2Grid}

\subsubsection{Description}

\texttt{modGrid2Grid} is not a class but a Fortan module which contains an array of \texttt{Vol2Vol} classes and subroutines to communicate with other languages. The module structure is depicted in Fig. \ref{fig:modGrid2GridStructure}. It has two static data which are input data array and \texttt{Vol2Vol} array not to have and update multiple \texttt{Vol2Vol} classes for efficiency and to save memory. When \texttt{Grid2Grid} is initialized from a program written in another language, it firstly check the input variables and return \texttt{HOSIndex}. If the same input variables are given, it returns the same \texttt{HOSIndex} and does not allocate nor initialise \texttt{Vol2Vol}. Other languages can distiguish different HOS wave theory by using the \texttt{HOSIndex}.
The subroutines \texttt{correct} and \texttt{get*} can be called with \texttt{HOSIndex}, position and time. 

The array size of \texttt{modGrid2Grid} is 100 by default (It can deal with 100 different HOS waves) but it does not consume much memory because classes of \texttt{Grid2Grid} are programmed with dynamic arrays and consequently only a few of static data are used. If more HOS wave theories are needed, the variable \texttt{nMaxVol2Vol} in \texttt{src/modGrid2Grid.f90} can be changed to deal with over than 100 different HOS wave theory. 

The functionality of \texttt{modGrid2Grid} is almost the same with \texttt{Vol2Vol}. 

\vspace{0.5cm}

{
	\begin{figure} [H]
		\centering
		\includegraphics[scale=0.72]{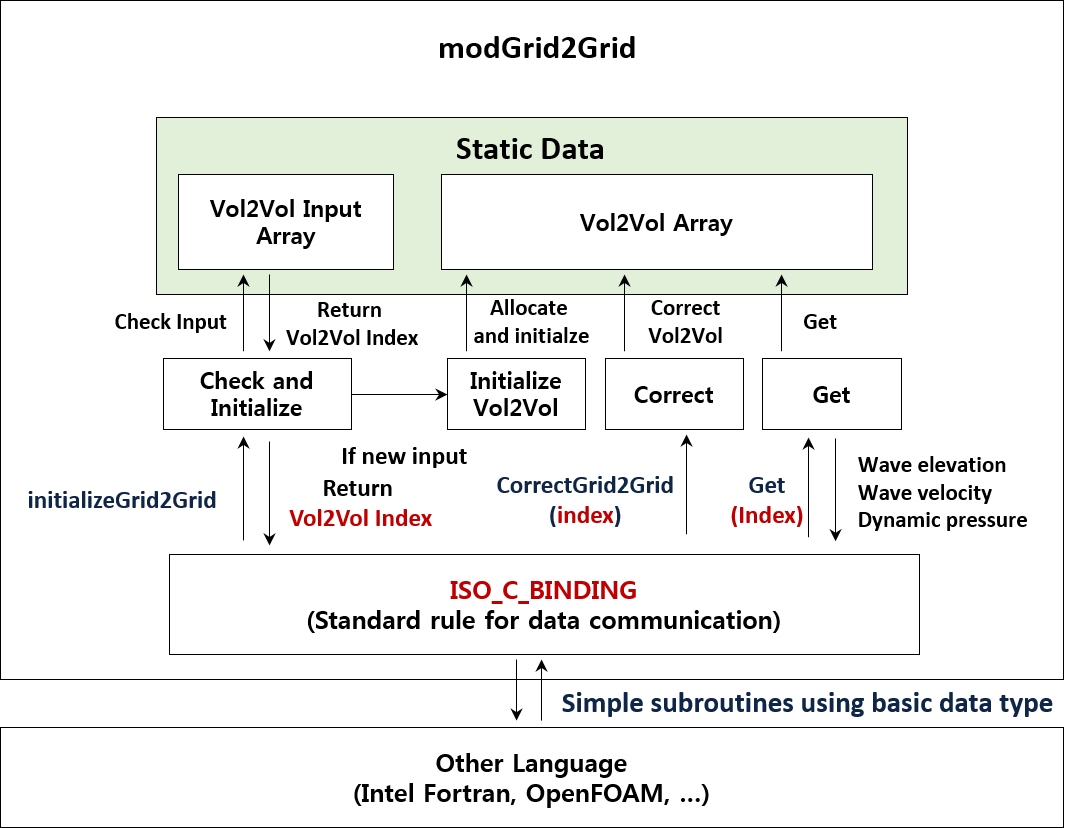}
		\vspace{0.5cm}
		\caption{\texttt{Grid2Grid} Module structure}
		\label{fig:modGrid2GridStructure}
	\end{figure}
}

\subsubsection{Subroutines}	

The interface between other languages with \texttt{modGrid2Grid} will be detailed in Chap. \ref{chap:interfaceLang}. 

Subroutines in \texttt{modGrid2Grid} follows \texttt{ISO\_C\_BINDING}. 

\vspace{0.5cm}

Static subroutines : 

\vspace{0.5cm}

-- Initialise Grid2Grid

\hspace{0.5 cm} $\circ$ \texttt{\_\_modgrid2grid\_MOD\_initializegrid2grid}
\vspace{0.2cm}

-- Correct Grid2Grid

\hspace{0.5 cm} $\circ$ \texttt{\_\_modgrid2grid\_MOD\_correctgrid2grid}
\vspace{0.2cm}

-- Get wave elevation

\hspace{0.5 cm} $\circ$ \texttt{\_\_modgrid2grid\_MOD\_gethoseta}
\vspace{0.2cm}

-- Get wave velocity

\hspace{0.5 cm} $\circ$ \texttt{\_\_modgrid2grid\_MOD\_gethosu}
\vspace{0.2cm}

-- Get dynamic pressure

\hspace{0.5 cm} $\circ$ \texttt{\_\_modgrid2grid\_MOD\_gethospd}
\vspace{0.2cm}

-- Get flow information 

\hspace{0.5 cm} $\circ$ \texttt{\_\_modgrid2grid\_MOD\_gethosflow}
\vspace{0.2cm}

-- Get HOS simulation end time

\hspace{0.5 cm} $\circ$ \texttt{\_\_modgrid2grid\_MOD\_gethosendtime}
\vspace{0.2cm}

-- Get HOS water depth

\hspace{0.5 cm} $\circ$ \texttt{\_\_modgrid2grid\_MOD\_gethoswaterdepth}
\vspace{0.2cm}

-- Get logical data indicating HOS wave theory is initialized

\hspace{0.5 cm} $\circ$ \texttt{\_\_modgrid2grid\_MOD\_isgrid2gridinitialized}

\pagebreak
	\subsection{PostGrid2Grid}
	
	\label{chap:postGrid2Grid}

	\subsubsection{Description}
	
	\texttt{PostGrid2Grid} is a HOS post-processing class. It generates 3D VTK files of wave fields for visualization and wave elevation time series computed from \texttt{Vol2Vol} class. Wave fields at desired simulation time and spatial domain and wave elevation time series can be re-generated at some provided wave probes position. 

	\texttt{PostGrid2Grid} algorithm is depicted in Fig. \ref{fig:postGrid2GridAlgorighm}. \texttt{PostGrid2Grid} is initialised with input file. The input file \texttt{postGrid2Grid.inp} contains HOS grid information and post processing information. \texttt{PostGrid2Grid} first reads and checks the input file and then build 3D visualization grid and wave probes. \texttt{Vol2Vol} class is also initialised. The subroutine \texttt{doPostProcessing} do time loop of \texttt{correct}. Subroutine \texttt{correct} first corrects the \texttt{Vol2Vol} class and gets the wave fields. If the grid option is set to no air mesh, 3D grid is fitted to wave elevation. It writes the results on files (3D VTK file and wave elevation time series). 
	
	\vspace{0.2cm}
	
	{
		\begin{figure} [H]
			\centering
			\includegraphics[scale=0.78]{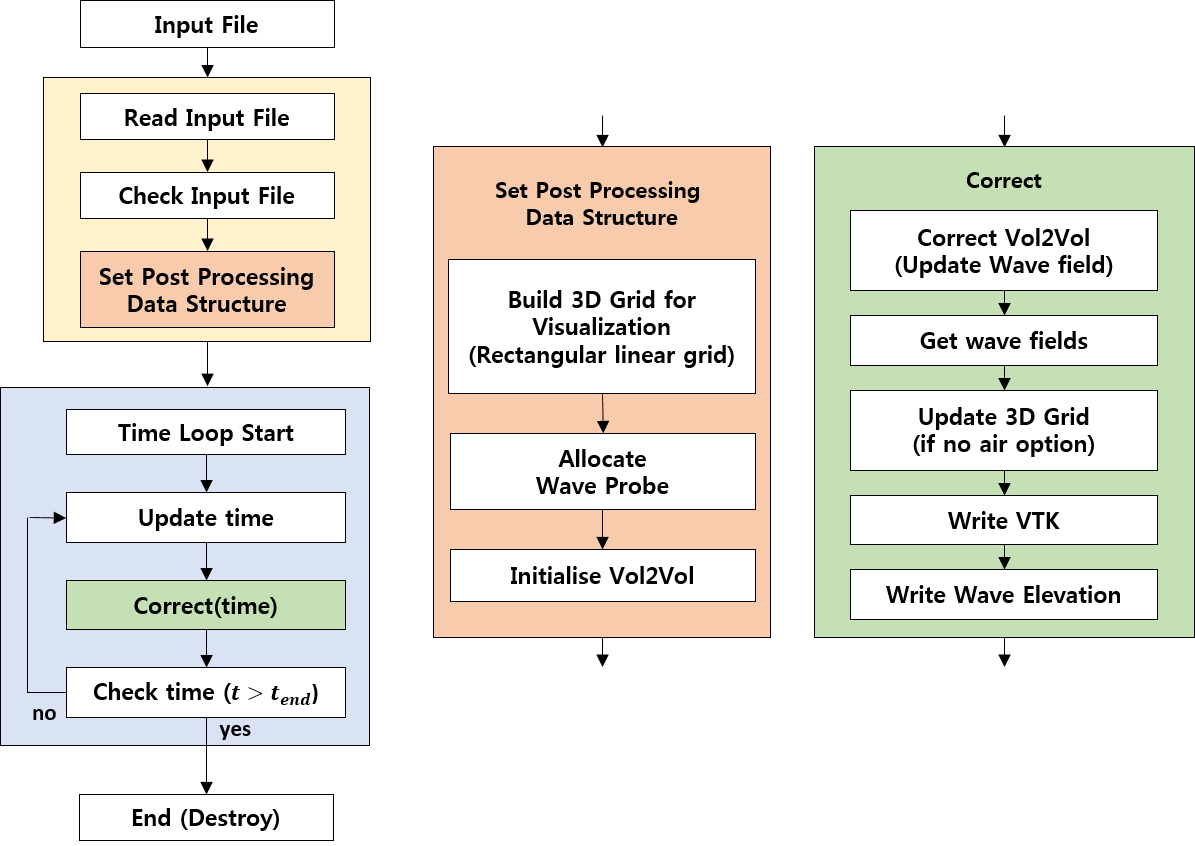}
			\vspace{0.5cm}
			\caption{\texttt{PostGrid2Grid} Algorithm}
			\label{fig:postGrid2GridAlgorighm}
		\end{figure}
	}	
	
	\pagebreak
	\subsubsection{Class(Type)}	
	
	\textbf{Class} : \texttt{PostGrid2Grid}
	
	\hspace{0.5 cm} -- Data :
	
	\hspace{1.0 cm} $\circ$ \texttt{hosVol2Vol\_} : \texttt{Vol2Vol} class
	
	\hspace{1.0 cm} $\circ$ \texttt{rectLGrid\_} : Rectangular linear grid for 3D wave fields (VTK output)
	
	\hspace{1.0 cm} $\circ$ \texttt{waveProbe\_(:)} : Wave probe 
	
	\vspace{0.5cm}
	
	\hspace{0.5 cm} -- Functionality (Public) :
	
	\hspace{1.0 cm} $\circ$ \texttt{initialize} : Initialise PostGrid2Grid class
	
	\hspace{1.0 cm} $\circ$ \texttt{correct} : Correct \texttt{Vol2Vol}, \texttt{rectLGrid\_} and \texttt{waveProbe\_} and write output
	
	\hspace{1.0 cm} $\circ$ \texttt{writeVTK} : Write 3D wave fields in VTK format
	
	\hspace{1.0 cm} $\circ$ \texttt{doPostProcessing} : Do post processing 
	
	\hspace{1.0 cm} $\circ$ \texttt{destroy} : Destuctor of PostGrid2Grid class
	
	\vspace{0.5cm}
	
	\hspace{0.5 cm} -- Functionality (Private) :
	
	\hspace{1.0 cm} $\circ$ \texttt{readPostG2GInputFile} : Read \texttt{PostGrid2Grid} input file
	
	\hspace{1.0 cm} $\circ$ \texttt{checkPostG2GParameter} : Check \texttt{PostGrid2Grid} input file
	
	\hspace{1.0 cm} $\circ$ \texttt{writeVTKtotalASCII} : Write wave fields (Including air domain)
	
	\hspace{1.0 cm} $\circ$ \texttt{writeVTKnoAirASCII} : Write wave fields (Grid is fitted to wave elevation)
	
	\hspace{1.0 cm} $\circ$ \texttt{writeWaveProbe} : Write wave elevation time series
	
	\pagebreak
	\subsubsection{Input File of PostGrid2Grid}
	
	\texttt{PostGrid2Grid} needs input file. The input file name is \texttt{postGrid2Grid.inp}. The input is recognized by \texttt{keyword}. Input file has free format. \texttt{keyword} can be located at any line of file but \texttt{keyword} should be located as first word at line. If special character is added on character, it is recognized as comment. The input keyword is following. 
	
	\vspace{0.5cm}
	-- HOS Type (\texttt{solver}) and Result File (\texttt{hosFile})
	
	\begin{lstlisting}[language=bash]
	
	### Select Solver (Ocean or NWT) ----------------------------------- #
	
	solver		Ocean
	
	### hosFile Path --------------------------------------------------- #
	
	hosFile 	modes_HOS_SWENSE.dat
	\end{lstlisting}
	
	-- Post Processing Time (\texttt{startTime}, \texttt{endTime}, \texttt{dt})
		
	\begin{lstlisting}[language=bash]
	
	### Post Processing Time ------------------------------------------- #
	
	startTime		2712.0
	endTime			2812.0
	dt	     		0.1	
	\end{lstlisting}
	
	-- Write VTK Option (\texttt{writeVTK}) and Air Meshing Option  (\texttt{airMesh})
	
	\begin{lstlisting}[language=bash]
	
	### Write 3D VTK File (true or false) ------------------------------ #
	
	writeVTK		true	
	
	### Air Meshing (true or false) ------------------------------------ #
	
	airMesh			 true
	
	#	If airMesh is true or yes, grid will be constructed up to zMax.
	#				     is false or no, mesh will be constructed up to wave 
	#					   elevation.
	\end{lstlisting}
	
	\pagebreak
	
	-- 3D Output Domain Size (\texttt{xMin}, \texttt{xMax}, \texttt{yMin}, \texttt{yMax}, \texttt{zMin}, \texttt{zMax})
	
	\begin{lstlisting}[language=bash]
	
	### 3D Output Domain Size ------------------------------------------ #
	
	xMin		1200
	xMax		2000
	
	yMin		1200
	yMax		2000
	
	zMin	-100.0
	zMax		50.0	

	# 	zMin and zMax are used to construct Surf2Vol HOS Grid.
	\end{lstlisting}
	
	-- Number of Grid (\texttt{nX}, \texttt{nY}, \texttt{nZmin}, \texttt{nZmax})
	
	\begin{lstlisting}[language=bash]
	
	### Number of Mesh for 3D Output ----------------------------------- #
	
	nX          500
	nY          500
	
	nZmin       100
	nZmax        60
	
	# 	nZmin and nZmax are used to construct Surf2Vol HOS Grid.
	\end{lstlisting}
	
	-- Vertical meshing scheme (\texttt{zMesh})
	
	\begin{lstlisting}[language=bash]
	
	### Z meshing scheme ----------------------------------------------- #
	
	zMesh       meshRatio    3.0    3.0
	
	#	Meshing Scheme
	#	
	#	uniform 	: uniform grid 
	#	sine			: sine spaced grid (densed grid near to z=0)
	#	meshRatio	: grid with constant geometric ratio. Minimum dz located  
	#					    at z=0. (given two ratio is dz_max/dz_min)
	#
	#	meshRatio needs two ratios (ratio for z <= 0  and ratio for z > 0)
	#
	\end{lstlisting}
	
	\pagebreak
	
	-- Wave probe write option (\texttt{writeWaveProbe}) and Output file path (\texttt{waveProbeFile})
	
	\begin{lstlisting}[language=bash]
	
	### Wave probe write option (true or false) ------------------------ #
	
	writeWaveProbe  true
	
	### Wave Probe File path ------------------------------------------- #
	
	waveProbeFile   waveElevation.dat

	#	If waveProbeFile is not given in input file. 
	#	Default iutput file name "waveElevation.dat" is used. 
	\end{lstlisting}
	
	-- Wave probes
	
	\begin{lstlisting}[language=bash]
	
	### Wave probe Input Format ---------------------------------------- #
	#
	#   There should be no blank line after nWaveProbe and probe data.
	#   If blank line is given, last wave probes will be discarded.
	#
	#		There should be no blank line. 
	#
	#		## Format
	#
	# 		nWaveProbe  nProbes
	# 		probe1name(option)     xPos1    yPos1
	# 		probe2name             xPos2    yPos2
	# 		...
	# 		probeNname             xPosN    yPosN
	#
	# ------------------------------------------------------------------ #
	
	### Wave Probe Input ----------------------------------------------- #
	
	nWaveProbe  5
	wp1     0.0     0.0
	wp2     1.0     2.0
	wp3     2.0     3.0
	wp4     4.0     5.0
	6.0 		7.0	

	#	probeName		xPosition		yPosition
	#
	#	probeName is optional.
	\end{lstlisting}	
	
	\pagebreak
	\subsubsection{How to use}
	
	Fortran subroutine for \texttt{postGrid2Grid} is given as example.
	
	\vspace{0.5cm}
	
	\begin{lstlisting}[language={[95]Fortran}]	

	!! Subroutine to run PostGrid2Grid -----------------------------------
	Subroutine runPostGrid2Grid(inputFileName)
	!! -------------------------------------------------------------------
	Use modPostGrid2Grid					!! Use PostGrid2Grid Module
	!! -------------------------------------------------------------------
	Implicit none	
	Character(Len = * ),intent(in) :: inputFileName		!! postGrid2Grid.inp	
	Type(typPostGrid2Grid)         :: postG2G				!! postGrid2Grid Class
	!! -------------------------------------------------------------------	
	
	!! Initialize PostGrid2Grid with Input File	
	Call postG2G%initialize(inputFileName)
	
	!! Do Post Processing
	Call postG2G%doPostProcessing()
	
	!! Destroy
	Call postG2G%destroy
	
	!! -------------------------------------------------------------------
	End Subroutine
	!! -------------------------------------------------------------------	
	\end{lstlisting}

\pagebreak
	\section{Installation}
	
		\subsection{Pre-Install}
		
		\subsubsection{FFTW Install}
		
		\label{chap:FFTWInstall}
		
		\texttt{Grid2Grid} needs fast Fourier transform (FFT) library. \texttt{FFTW} is GNU licenced FFT library. The installation order is the following:
		
		\vspace{0.5cm}
		
		\textbf{1. Download FFTW library}
		
		At the FFTW website (http://www.fftw.org/download.html), latest version of FFTW is available. Download \texttt{FFTW} library. 
		
		\vspace{0.5cm}
		{
			\begin{figure} [H]
				\centering
				\begin{tcolorbox}[standard jigsaw,opacityback=0]
				\includegraphics[scale=0.72]{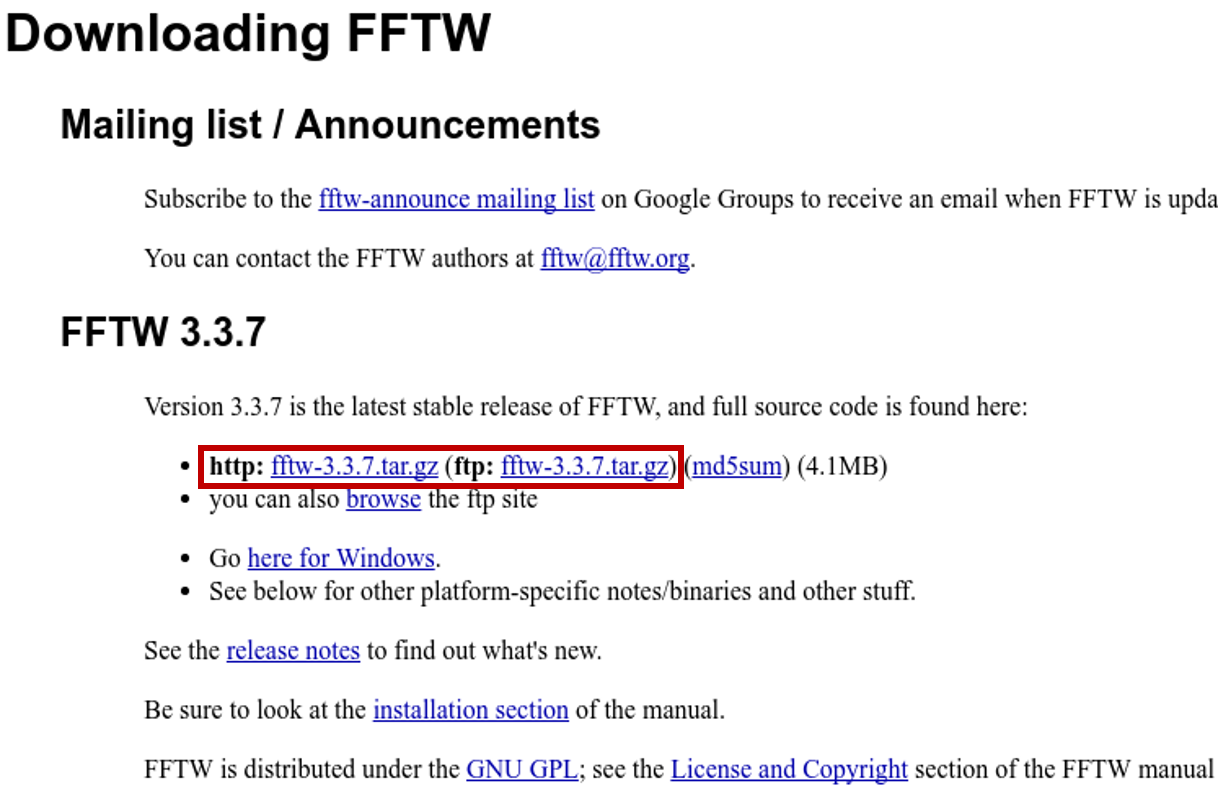}
				\end{tcolorbox}
				\vspace{0.3cm}
				\caption{Download \texttt{FFTW library}}			
				\label{fig:vol2volStructure}
			\end{figure}
		}
		
		\pagebreak
		
		\textbf{2. Extract FFTW library}

		\begin{lstlisting}[language=bash]
		
		$	tar -xvf fftw-3.3.7.tar.gz
		\end{lstlisting}
		
		\vspace{0.5cm}
		\textbf{3. Compile FFTW library}
		
		\begin{lstlisting}[language=bash]
		
		$	cd fftw-3.3.7/
		$
		$	export FFTW_PATH=$PWD
		$
		$	./configure --prefix=$FFTW_PATH		
		$
		$	make CFLAGS='-fPIC'
		$
		$	make install
		\end{lstlisting}

		\vspace{0.5cm}
		\textbf{4. Make soft link of FFTW library }
		
		If user has super user authority
		
		\begin{lstlisting}[language=bash]
		
		$	sudo ln -s $FFTW_PATH/lib/libfftw3.a  /usr/local/lib/libfftw3.a		
		\end{lstlisting}
				
		\vspace{0.5cm}
		
		If user has no super user authority, third party path in \texttt{makefile} of \texttt{Grid2Grid} can be changed manually. If FFTW library (\texttt{libfftw3.a}) locates at \texttt{/home/lib/libfftw3.a}, then \texttt{makefile} of \texttt{Grid2Grid} can be changed as:

		\begin{lstlisting}[language=bash]
		
		DTHRDLIB=/home/lib/
		\end{lstlisting}

		\pagebreak
		\subsubsection{HOS Ocean and NWT (Optional)}
		
		\texttt{Grid2Grid} needs the result file of HOS solver. Installation of HOS is the following: 
		
		\vspace{0.2cm}
		\textbf{1. Install FFTW}
		
		See Chapter \ref{chap:FFTWInstall}
		
		\vspace{0.2cm}
		\textbf{2. Install makedepf90}
		
		\begin{lstlisting}[language=bash]		
		
		$	sudo apt-get install makedepf90		
		\end{lstlisting}
		
		\vspace{0.2cm}
		\textbf{3. Install LAPACK}
		
		\begin{lstlisting}[language=bash]
		
		$	sudo apt-get install liblapack-dev liblapack-doc-man liblapack-doc liblapack-pic liblapack3 liblapack-test liblapack3gf liblapacke liblapacke-dev
		\end{lstlisting}
		
		\vspace{0.2cm}
		\textbf{4. Download HOS Ocean and NWT}
		
		\begin{lstlisting}[language=bash]
		
		$	# Path to desired installation path
		$
		$	cd $HOS_INSTALLATION_PATH
		$
		$	git clone https://github.com/LHEEA/HOS-ocean.git
		$
		$	git clone https://github.com/LHEEA/HOS-NWT.git		
		\end{lstlisting}
		
		\vspace{0.2cm}
		\textbf{5. Change shared library path defined in makefile}
		
		Open \texttt{makefile} of HOS Ocean and NWT and change as following : 
		
		\begin{lstlisting}[language=bash]

		#LINKLIB = $(LIBDIR)libfftw3.a $(LIBDIR)liblapack.a $(LIBDIR)librefblas.a
		
		LINKLIB = $(LIBDIR)libfftw3.a -llapack
		\end{lstlisting}
		
		If FFTW s not installed at \texttt{\$(LIBDIR)} in \texttt{makefile}, it can be changed as an example of previous chapter. 
		
		\pagebreak
		
		\vspace{0.2cm}
		\textbf{6. Comple HOS NWT and HOS Ocean}
		
		\begin{lstlisting}[language=bash]

		$	cd HOS-ocean/
		$
		$	make
		$	
		$	cd ../HOS-NWT/
		$
		$	make
		$
		$	cd ..		
		\end{lstlisting}
		
		\vspace{0.2cm}
		\textbf{7. Check executable is generated}
		
		\begin{lstlisting}[language=bash]
		
		$	# Check HOS NWT
		$	ls HOS-NWT/bin/HOS-NWT
		$
		$	# Check HOS Ocean
		$	ls HOS-ocean/bin/HOS-ocean
		\end{lstlisting}
		
		\vspace{0.2cm}
		\textbf{8. Make soft link (optional)}
		
		\begin{lstlisting}[language=bash]

		$	# Make Soft Link (Optional)
		$
		$	export HOS_PATH=$PWD
		$
		$	sudo ln -s $HOS_PATH/bin/HOS-NWT		/usr/bin/HOS-NWT
		$
		$	sudo ln -s $HOS_PATH/bin/HOS-ocean	/usr/bin/HOS-ocean
		\end{lstlisting}

		\pagebreak
		\subsection{Grid2Grid Installation}
		
		\label{chap:Grid2GridInstall}
		
		\subsubsection{Download Grid2Grid}
		
		\textbf{Download Grid2Grid}
		
		\begin{lstlisting}[language=bash]
		
		$	# Path to desired installation path
		$
		$	cd $HOS_INSTALLATION_PATH
		$
		$	git clone https://github.com/LHEEA/Grid2Grid.git
		\end{lstlisting}
		
		\vspace{0.2cm}
		
		\textbf{Set FFTW Path (FFTW library is not installed on \texttt{/usr/local/lib/}}
		
		Change \texttt{DTHRDLIB} in \texttt{makefile} at FFTW library path.
		
		\begin{lstlisting}[language=bash]
		
		DTHRDLIB=$FFTW_LIBRARY_PATH
		
		# Example
		#	DTHRDLIB=/home/lib/			if "libfftw3.a" exists at /home/lib/
		\end{lstlisting}
		
		\subsubsection{Compile PostGrid2Grid}
		
		\begin{lstlisting}[language=bash]										

		$	cd Grid2Grid
		$
		$	make 
		\end{lstlisting}
		
		\subsubsection{Compile Grid2Grid Shared Library}
		
		\textbf{Compile \texttt{libGrid2Grid.so} in \texttt{Grid2Grid/obj/}}
		
		\begin{lstlisting}[language=bash]
		
		$	make createlib
		\end{lstlisting}		
	
		\vspace{0.2cm}

		\textbf{Compile \texttt{libGrid2Grid.so} in \texttt{\$FOAM\_USER\_LIBBIN}}
		
		If OpenFOAM is installed, \texttt{libGrid2Grid.so} can be compiled at \texttt{\$FOAM\_USER\_LIBBIN}. If OpenFOAM environment is called, following make rule can be used directly. 
		
		\begin{lstlisting}[language=bash]
		
		$	make createOFlib
		\end{lstlisting}

\pagebreak
	\section{Interface}
	\label{chap:interfaceLang}

		\subsection{GNU and Intel Fortran}

		An interface example fortran program is included in \texttt{interface/fortGrid2Grid}. To communicate with \texttt{Grid2Grid}, the fortran script \texttt{modCommG2G.f90} in \texttt{interface/}\texttt{fortGrid2Grid} is needed. It is a communication module with \texttt{libGrid2Grid.so}. The fortran interface example program is following :

		\begin{lstlisting}[language={[95]Fortran}]

	!! Program Start ----------------------------------------------------
	Program Main
	!! ------------------------------------------------------------------
	use modCommG2G			!! Use Communication Module
	!! ------------------------------------------------------------------
	Implicit None
	!! Variables --------------------------------------------------------
	Integer,Parameter      :: nChar = 300			!! Default Character Length
	Character(len = nChar) :: grid2gridPath		!! libGrid2Grid.so Path

	integer                :: hosIndex				!! HOS Index
	Character(len = nChar) :: hosSolver				!! HOS Solver (Ocean or NWT)
	Character(len = nChar) :: hosFileName			!! HOS Result File Path

	Double precision       :: zMin, zMax			!! Surf2Vol Domain
	integer                :: nZmin, nZmax		!! Number of vertical grid
	Double precision       :: zMinRatio, zMaxRatio	!! Grading ratio (=3)

	Double precision       :: t, dt			!! Simulation Time, dt
	Double precision       :: x, y, z		!! Computation Point
	Double precision       :: eta, u, v, w, pd	!! HOS Wave Information
	!! Dummy variables --------------------------------------------------
	integer                :: it				!! Dummy time loop integer

	!! Program Body -----------------------------------------------------

	!!!... Write Program Start
	write(*,*) "Test program (Connect to Fortran) to use Grid2Grid shared library"

	!!!... Set libGrid2Grid.so path.
	!!!    It is recommended to use absolute path
	! grid2gridPath = "/usr/lib/libGrid2Grid.so"	(if soft link is made)
	grid2gridPath = "../../obj/libGrid2Grid.so"

	!!!... Load libGrid2Grid.so and connect subroutines
	Call callGrid2Grid(grid2gridPath)

	!!!... Declare HOS Index
	hosIndex = -1

	!!!... Set HOS Type (Ocean or NWT)
	hosSolver = "NWT"

	!!!... Set HOS Result file Path
	hosFileName = "modes_HOS_SWENSE.dat"

	!!!... Set HOS Surf2Vol Domain and Vertical Grid
	zMin = -0.6d0; 				zMax =  0.6d0
	nZmin = 50; 					nZmax = 50
	zMinRatio = 3.d0; 		zMaxRatio = 3.d0

	!!... Initialize Grid2Grid and Get HOS Index
	Call initializeGrid2Grid(hosSolver, hosFileName, zMin, zMax, nZmin, nZmax, zMinRatio, zMaxRatio, hosIndex)

	!! Time Information
	t  = 0.0d0; 		dt = 0.1d0

	!! Given Point
	x = 0.5d0; 			y = 0.5d0; 			z = -0.5d0

	!! Time Loop
	do it = 1,10

		!! Correct HOS Vol2VOl for given time
		Call correctGrid2Grid(hosIndex, t)

		!! Get Wave Elevation
		Call getHOSeta(hosIndex, x, y , t, eta)

		!! Get Flow Velocity
		Call getHOSU(hosIndex, x, y, z, t, u, v ,w)

		!! Get Dynamic Pressure
		Call getHOSPd(hosIndex, x, y, z, t, pd)

		!! Write Flow Information
		write(*,*) t, eta, u, v, w, pd

		!! Time Update
		t = t + dt
	enddo

	!! Write End of Program
	write(*,*) "Test program (Connect to Fortran) is done ..."
	!! ------------------------------------------------------------------
	End Program
	!! ------------------------------------------------------------------
		\end{lstlisting}

		\pagebreak
		\subsection{OpenFOAM}

		An interface example OpenFOAM program is included in \texttt{interface/ofGrid2Grid}. The shared library \texttt{libGrid2Grid.so} should be compiled at \texttt{\$FOAM\_USER\_LIBBIN}. To check \texttt{libGrid2Grid.so} exists at \texttt{\$FOAM\_USER\_LIBBIN}, use following shell command :

		\begin{lstlisting}[language=bash]

		$	ls $FOAM_USER_LIBBIN/libGrid2Grid.so
		\end{lstlisting}

		If \texttt{libGrid2Grid.so} not exists, refer to Chapter \ref{chap:Grid2GridInstall}.

		To call shared library \texttt{libGrid2Grid.so} in \texttt{\$FOAM\_USER\_LIBBIN}, OpenFOAM compiling option is added at \texttt{Make/option}. Open \texttt{Make/option} and add following compiling option.

		\begin{lstlisting}[language=bash]

		EXE_LIBS = \
							...
							-lgfortran \
							-L$(FOAM_USER_LIBBIN) \
							-lGrid2Grid
		\end{lstlisting}

		OpenFOAM interface example program is given next page.

		\pagebreak

		\begin{lstlisting}[language=c++]

#include "fvCFD.H"

namespace Foam
{
	//- Grid2Grid Initial Character Length
	const int nCharGridGrid(300);

	//- Initialize Grid2Grid Class in Fortran
	//
	//  __modgrid2grid_MOD_initializegrid2grid
	//  (
	//      hosSolver,
	//      hosFileName,
	//      zMin,
	//      zMax,
	//      nZmin,
	//      nZmax,
	//      zMinRatio,
	//      zMaxRatio,
	//      hosIndex
	//  )
	//
	//    Input
	//      hosSolver            : "NWT" or "Ocean"
	//      hosFileName          : filePath of HOS mode result file
	//      zMin, zMax           : HOS grid zMin and zMax
	//      nZmin, nZmax         : HOS number of z grid
	//      zMinRatio, zMaxRatio : HOS z grid max/min ratio
	//
	//    Output
	//      hosIndex             : HOS Vol2Vol Index
	//
	extern "C" void __modgrid2grid_MOD_initializegrid2grid
	(
	const char[nCharGridGrid],
	const char[nCharGridGrid],
	const double*,
	const double*,
	const int*,
	const int*,
	const double*,
	const double*,
	int*
	);

	//- Correct Grid2Grid for given simulation Time
	//
	//  __modgrid2grid_MOD_correctgrid2grid(hosIndex, simulTime)
	//
	//    Input
	//      hosIndex   : HOS Vol2Vol Index
	//      simulTime  : Simulation Time
	//
	extern "C" void __modgrid2grid_MOD_correctgrid2grid
	(
	const int *,
	const double *
	);

	//- Get HOS Wave Elevation
	//
	//  __modgrid2grid_MOD_gethoseta(hosIndex, x, y, t, eta)
	//
	//    Input
	//      hosIndex : HOS Vol2Vol Index
	//      x, y, t  : (x and y) position and simulation Time (t)
	//
	//    Output
	//      eta      : wave elevation
	//
	extern "C" void __modgrid2grid_MOD_gethoseta
	(
	const int *,
	const double *,
	const double *,
	const double *,
	double *
	);

	//- Get HOS Flow Velocity
	//
	//  __modgrid2grid_MOD_gethosu(hosIndex, x, y, z, t, u, v, w)
	//
	//    Input
	//      hosIndex    : HOS Vol2Vol Index
	//      x, y, z, t  : (x, y, z) position and simulation Time (t)
	//
	//    Output
	//      u, v, w     : (x, y, z) - directional flow velocity
	//
	extern "C" void __modgrid2grid_MOD_gethosu
	(
	const int *,
	const double *,
	const double *,
	const double *,
	const double *,
	double *,
	double *,
	double *
	);
	//- Get HOS Dynamic Pressure
	//
	//  __modgrid2grid_MOD_gethospd(hosIndex, x, y, z, t, pd)
	//
	//    Input
	//      hosIndex    : HOS Vol2Vol Index
	//      x, y, z, t  : (x, y, z) position and simulation Time (t)
	//
	//    Output
	//      pd      : Dynamic Pressure p = -rho*d(phi)/dt-0.5*rho*|U*U|
	//
	extern "C" void __modgrid2grid_MOD_gethospd
	(
	const int *,
	const double *,
	const double *,
	const double * ,
	const double *,
	double *
	);

	//- Get HOS Wave Elevation, Flow Velocity and Dynamic Pressure
	//
	//  __modgrid2grid_MOD_gethosflow(hosIndex, x, y, z, t, eta, u, v, w, pd)
	//
	//    Input
	//      hosIndex    : HOS Vol2Vol Index
	//      x, y, z, t  : (x, y, z) position and simulation Time (t)
	//
	//    Output
	//      eta         : wave elevation
	//      u, v, w     : (x, y, z) - directional flow velocity
	//      pd          : Dynamic Pressure p = -rho * d(phi)/dt - 0.5 * rho * |U * U|
	//
	extern "C" void __modgrid2grid_MOD_gethosflow
	(
	const int *,
	const double *,
	const double *,
	const double * ,
	const double *,
	double *,
	double *,
	double * ,
	double *,
	double *
	);

}

// Main OpenFOAM Program Start
int main(int argc, char *argv[])
{

	// Write Program Start
	Info << "OpenFOAM Program Example to Call Grid2Grid (HOS Wrapper) in OpenFOAM" << endl;

	// Set HOS Solver Type
	const word HOSsolver_("NWT");
	const word HOSFileName_("./modes_HOS_SWENSE.dat");

	// Set File Name
	string strHOSSolver = string(HOSsolver_);
	string strHOSFileName = string(HOSFileName_);

	// Set HOS Solver Type
	const char *HOSsolver = strHOSSolver.c_str();

	// Set HOS Mode Result File Path
	const char *HOSfileName = strHOSFileName.c_str();

	// Set HOS Z Grid Information
	int indexHOS(-1);

	double zMin(-0.6), zMax(0.6);
	int nZmin(50), nZMax(50);

	double zMinRatio(3.0), zMaxRatio(3.0);

	// Initialize Grid2Grid
	__modgrid2grid_MOD_initializegrid2grid(HOSsolver, HOSfileName,
	&zMin, &zMax,
	&nZmin, &nZMax,
	&zMinRatio, &zMaxRatio, &indexHOS);

	Info << "HOS Label : " << indexHOS << endl;

	// Set Position
	double x(0.5), y(0.0), z(-0.5);

	// Define Flow Quantities
	double eta, u, v, w, pd;

	// Set Simulation Time and Time Difference
	double simulTime(0.0);
	double dt(0.1);



	// Time Loop
	for (int it = 0; it < 11; it++)
	{
		// Correct Grid2Grid
		__modgrid2grid_MOD_correctgrid2grid(&indexHOS, &simulTime);

		// Get Wave Eta
		__modgrid2grid_MOD_gethoseta(&indexHOS, &x, &y, &simulTime, &eta);

		// Get Flow Velocity
		__modgrid2grid_MOD_gethosu(&indexHOS, &x, &y, &z, &simulTime, &u, &v, &w);

		// Get Dynamic Pressure
		__modgrid2grid_MOD_gethospd(&indexHOS, &x, &y, &z, &simulTime, &pd);

		Info << " sumulTime : " << simulTime << endl;
		Info << "   eta     : " << eta << endl;
		Info << "   u, v, w : " << u << " " << v << " " << w << endl;
		Info << "   pd      : " << pd << nl << endl;

		// Get whole Information
		__modgrid2grid_MOD_gethosflow(&indexHOS, &x, &y, &z, &simulTime, &eta, &u, &v, &w, &pd);
		Info << "   eta     : " << eta << endl;
		Info << "   u, v, w : " << u << " " << v << " " << w << endl;
		Info << "   pd      : " << pd << nl << endl;

		// Time Update
		simulTime+=dt;
	}

	return 0;
}
		\end{lstlisting}

\pagebreak
	\section{Validation}
	
	\texttt{foamStar} is used to validate \texttt{Grid2Grid}. \texttt{foamStar} is developed by Bureau Veritas and shared with Ecole Centrale de Nantes and it can simulate nonlinear waves, seakeeping problem and also hydro-elasticity problem. It solves multiphase problem with Reynolds Averaged Navier-Stokes equations (RANS) with Volume of fraction (VOF). It is based on standard multiphase solver in OpenFOAM (\texttt{interDymFoam}). To generate waves, \texttt{foamStar} uses explicit blending scheme which blends computed flow values to target values with weight function. The blending function is given as equation \eqref{eq:blendingScheme}. Some details of \texttt{foamStar} are explained in \cite{sopheakThesis} and \cite{Charles2017}.
	
	{
		\begin{subequations}
			\label{eq:blendingScheme}
			\begin{align}		
				\text{U} 	 &= (1-w)\text{U}  + w\text{U}^{target} \\
				\alpha 		&= (1-w)\alpha + w\alpha^{target}
			\end{align}
		\end{subequations}
		\centering
		where            
		\begin{equation*}
		w = 1 - (1-w_{base})^{\chi} \qquad  \qquad \chi = \dfrac{\Delta \text{U} \Delta t}{\Delta x} \qquad \qquad  w_{base} = -2\xi^3 + 3\xi^2
		\end{equation*} 
	}

	By using \texttt{Grid2Grid}, the target values at blending zone can be replaced by the wave components computed from HOS wave theory. For the validation, considered HOS simulation condition is given in Table \ref{table:HOSValidataionCase}.
	
	\vspace{0.2cm}	
	{
		\begin{table}[H]
			\caption{HOS Wave condition for validation}
			\label{table:HOSValidataionCase}
			\centering
			\begin{tabular}{  C{2.7cm}   C{1.5cm}   C{1.8cm} C{1.8cm} C{1.8cm} C{1.8cm } }
				\hline \hline				
				\multirow{2}{*}{Wave Type} & \multirow{2}{*}{Value} & \multicolumn{2}{c}{HOS-Ocean} &	\multicolumn{2}{c}{HOS-NWT} \\
				\cline{3-6}
														& 	& 2D & 3D & 2D & 3D \\				
				\hline
				\multirow{2}{*}{Regular Wave} & $T$ [s]    & -	 & - &  0.702 &	0.702 \\
															& $H$ [m]  & -	 & - &  0.0431 &	0.0288 \\
				\hline
				\multirow{3}{*}{Irregular Waves} & $T_p$ [s] & 0.702	& 1.0 &  1.0 &	0.702 \\
															 & $H_s$[m] & 0.0288 & 0.10 &  0.05 &	0.0384 \\
															 & $\gamma$ [-] & 3.3	& 3.3 &  3.3 &	3.3  \\
				\hline
				\multirow{3}{*}{Extreme Waves} & $T_p$ [s] & -	& - &  17.5 &	- \\
															  & $H_s$[m] & - & - &  15.5 &	- \\
														      & $\gamma$ [-] & -	& - &  3.3 &	-  \\
				\hline\hline
			\end{tabular}
		\end{table}
	}
	
	\pagebreak
	
	\subsection{Simulation results}
	
	The wave elevation computed by using \texttt{foamStar} and HOS-Ocean are compared in Fig. \ref{fig:HOSOcean_WaveElevation}. 
	Small differences are observed but it is assumed to be caused by the resolution of the finite volume mesh which is not sufficient near the free surface. The VOF solver gives then those differences. Snapshots of HOS-Ocean wave fields are shown in Fig. \ref{fig:HOSOcean_WaveElevation}. 

	The results of HOS-NWT for a two dimensional case is shown in Fig. \ref{fig:HOSNWT_2DWaveElevation}. And a simulation snapshot is shown in Fig. \ref{fig:HOSNWT_2DSnapshot}. Three dimensional waves with HOS-NWT are also given in Figs. \ref{fig:HOSNWT_3DWaveElevation} and \ref{fig:HOSNWT_3DSnapshot}. 
	
	\vspace{0.5cm}
	
	{
		\begin{figure} [H]
			\centering
			\subfloat[][ HOS-Ocean 2D Irregular Waves]
			{\includegraphics[scale=0.51, trim=1.2cm 0 0 0]{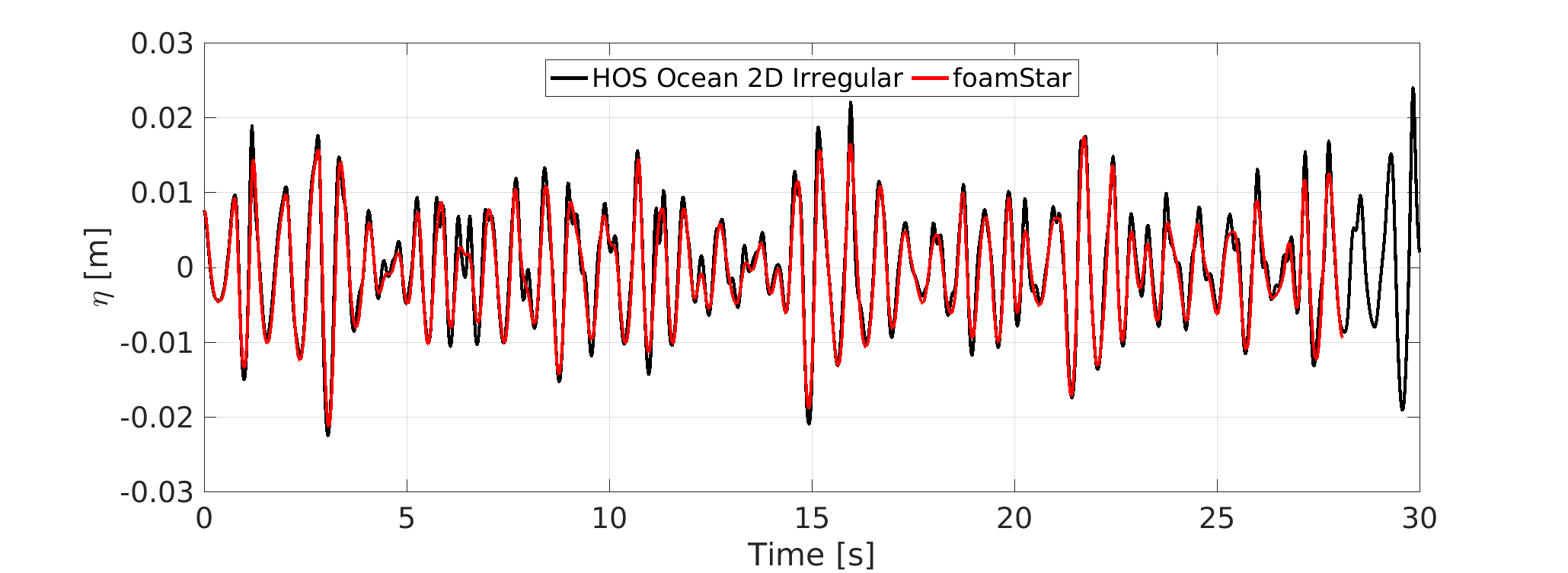}} \\
			\subfloat[][ HOS-Ocean 3D Irregular Waves]
			{\includegraphics[scale=0.51, trim=1.2cm 0 0 0]{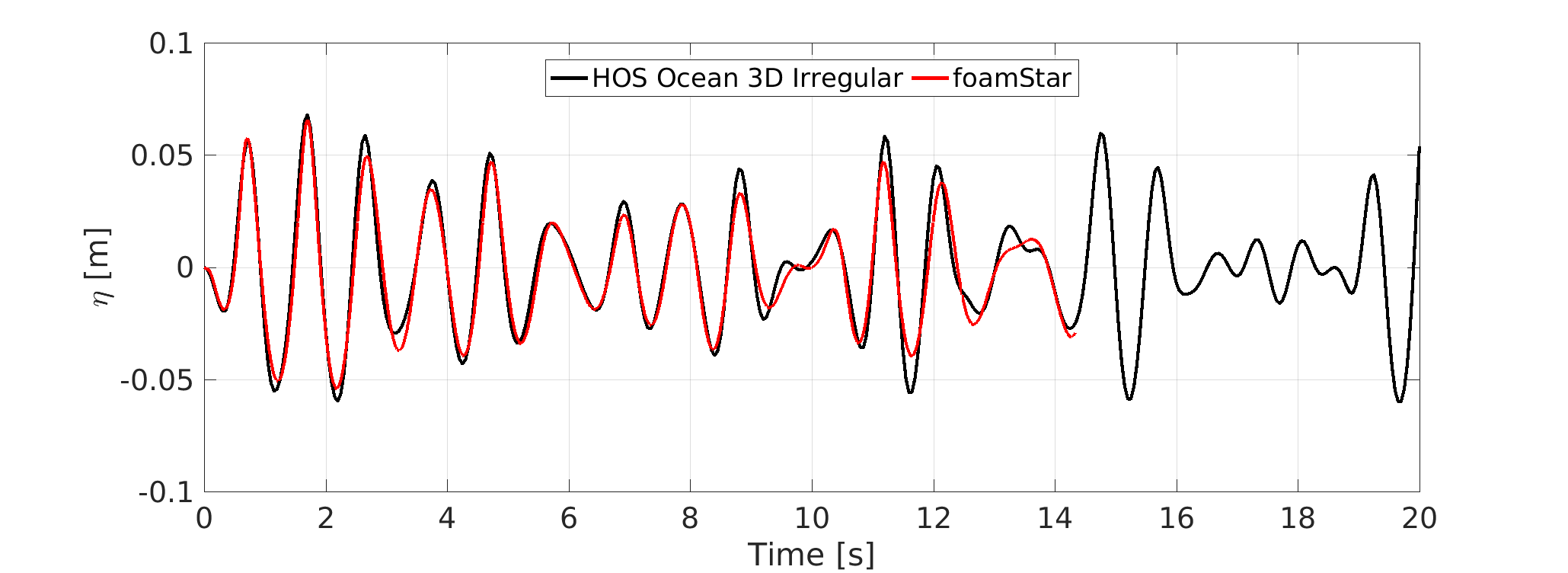}}
			\vspace{0.5cm}
			\caption{Comparison of HOS-Ocean wave elevation}
			\label{fig:HOSOcean_WaveElevation}
		\end{figure}
	}
	
	{
	\begin{figure} [H]
		\centering
		\subfloat[][ HOS-Ocean 2D Irregular Waves] 
		{\includegraphics[scale=1.25]{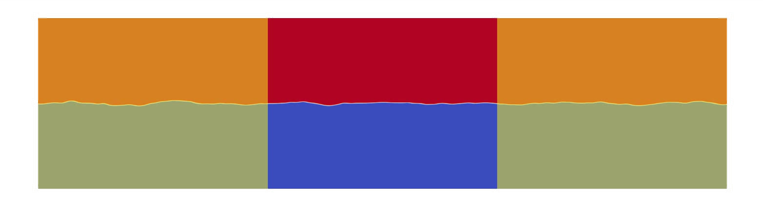}} \\
		\vspace{0.5cm}
		\subfloat[][ HOS-Ocean 3D Irregular Waves]
		{\includegraphics[scale=1.4]{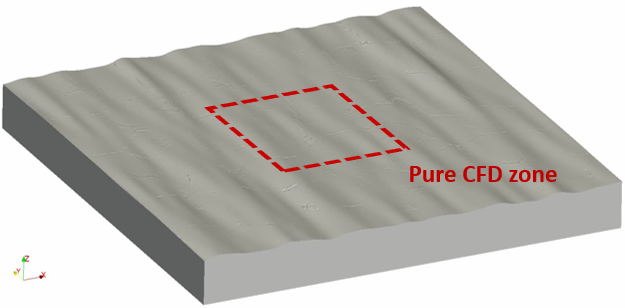}}
		\vspace{0.5cm}
		\caption{Snapshot of HOS-Ocean wave fields by \texttt{foamStar}}
		\label{fig:HOSOcean_Snapshot}
	\end{figure}
	}

	\pagebreak
	
	{
		\begin{figure} [H]
			\centering
			\subfloat[][ HOS-NWT 2D Regular Waves]
			{\includegraphics[scale=0.42, trim=1.2cm 0 0 0]{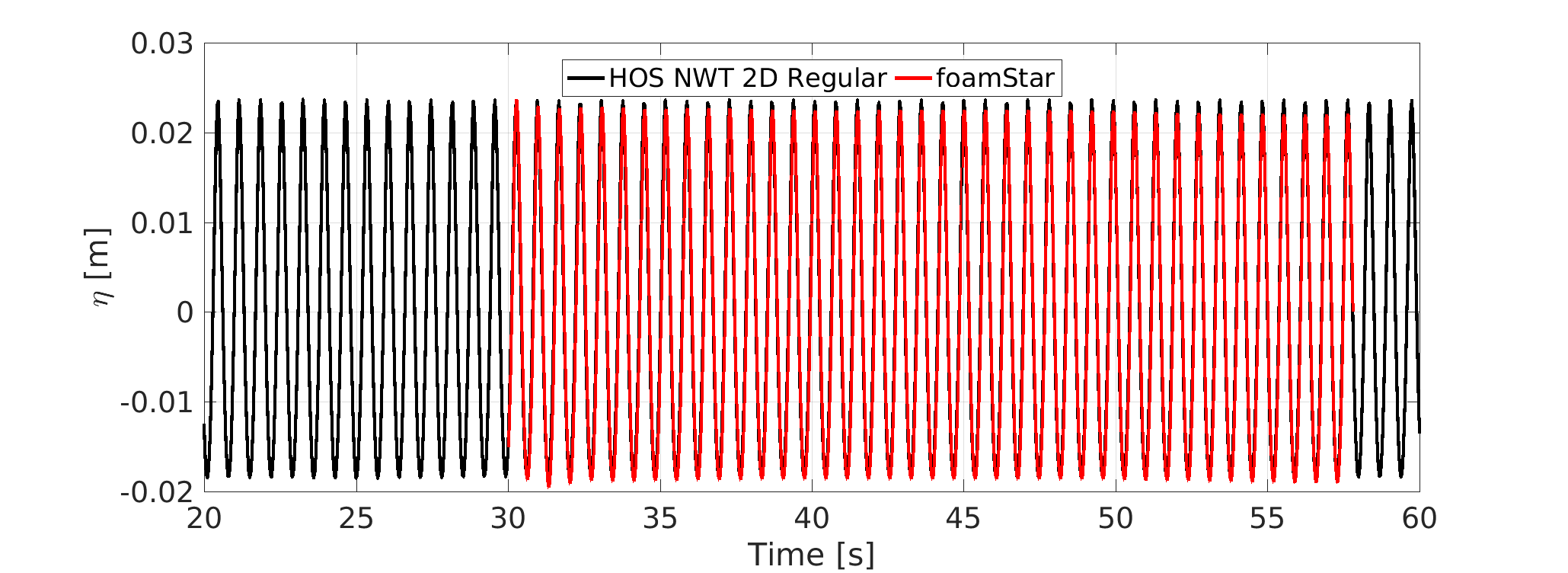}} \\
			\subfloat[][ HOS-NWT 2D Irregular Waves]
			{\includegraphics[scale=0.42, trim=1.2cm 0 0 1cm]{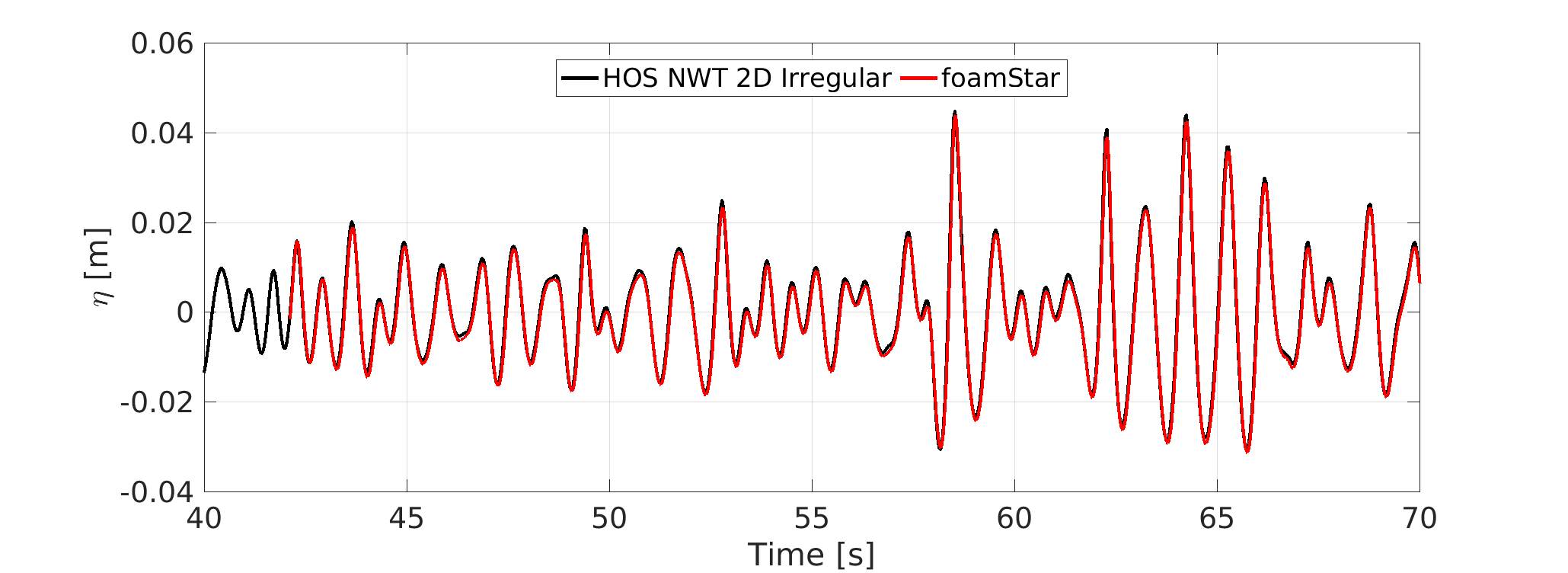}}
			\vspace{0.2cm}
			\caption{Comparison of HOS-NWT 2D wave elevation}
			\label{fig:HOSNWT_2DWaveElevation}
		\end{figure}
	}

	{
		\begin{figure} [H]
			\centering
			\subfloat[][ HOS-NWT 2D Irregular Waves] 
			{\includegraphics[scale=1.]{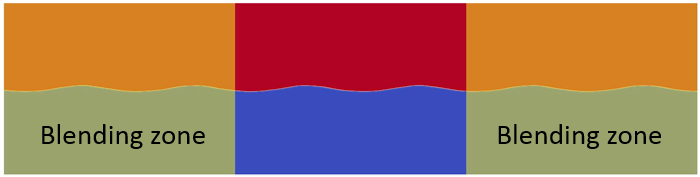}} \\
			\vspace{0.2cm}
			\subfloat[][ HOS-NWT 2D Irregular Waves]
			{\includegraphics[scale=1.]{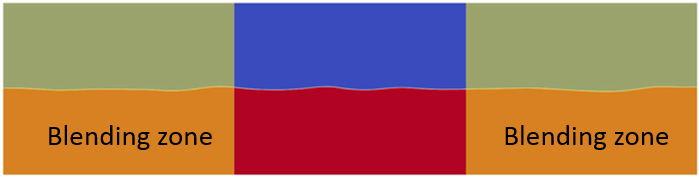}}
			\vspace{0.2cm}
			\caption{Snapshot of HOS-NWT 2D wave fields by \texttt{foamStar}}
			\label{fig:HOSNWT_2DSnapshot}
		\end{figure}
	}

	\pagebreak
	
	{
		\begin{figure} [H]
			\centering
			\subfloat[][ HOS-NWT 3D Regular Waves]
			{\includegraphics[scale=0.48, trim=1.2cm 0 0 0]{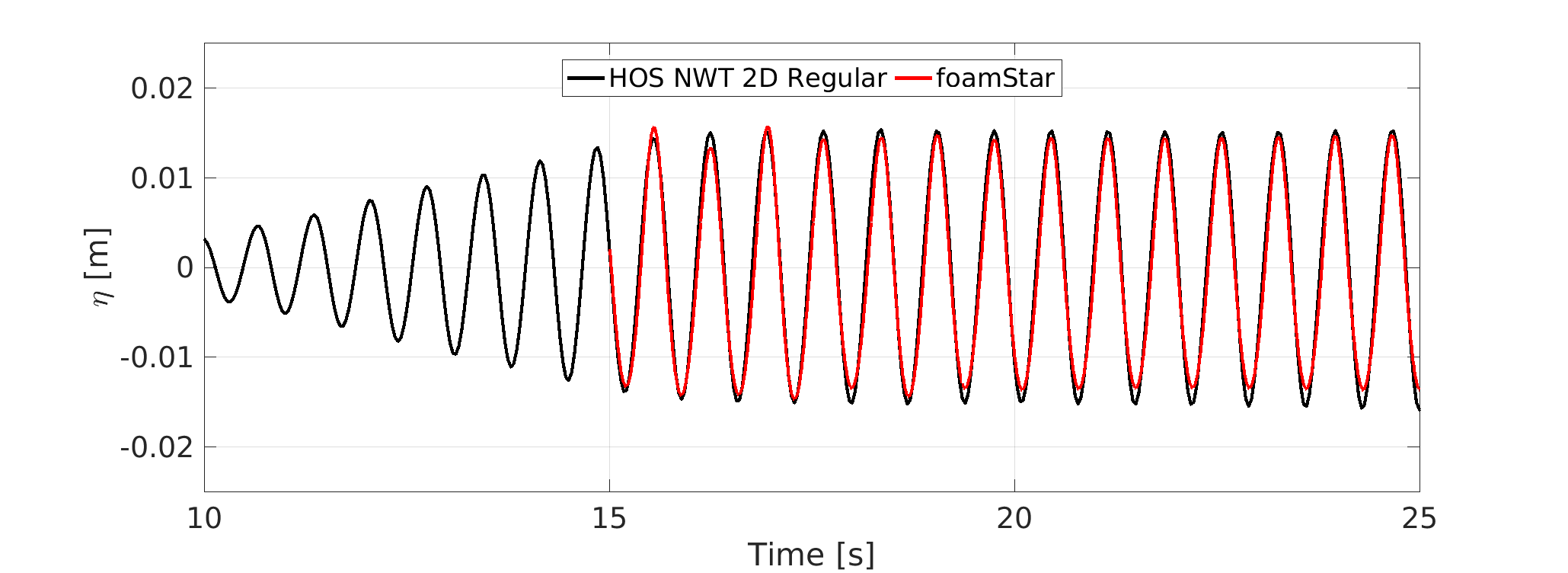}} \\
			\subfloat[][ HOS-NWT 3D Irregular Waves]
			{\includegraphics[scale=0.48, trim=1.2cm 0 0 1cm]{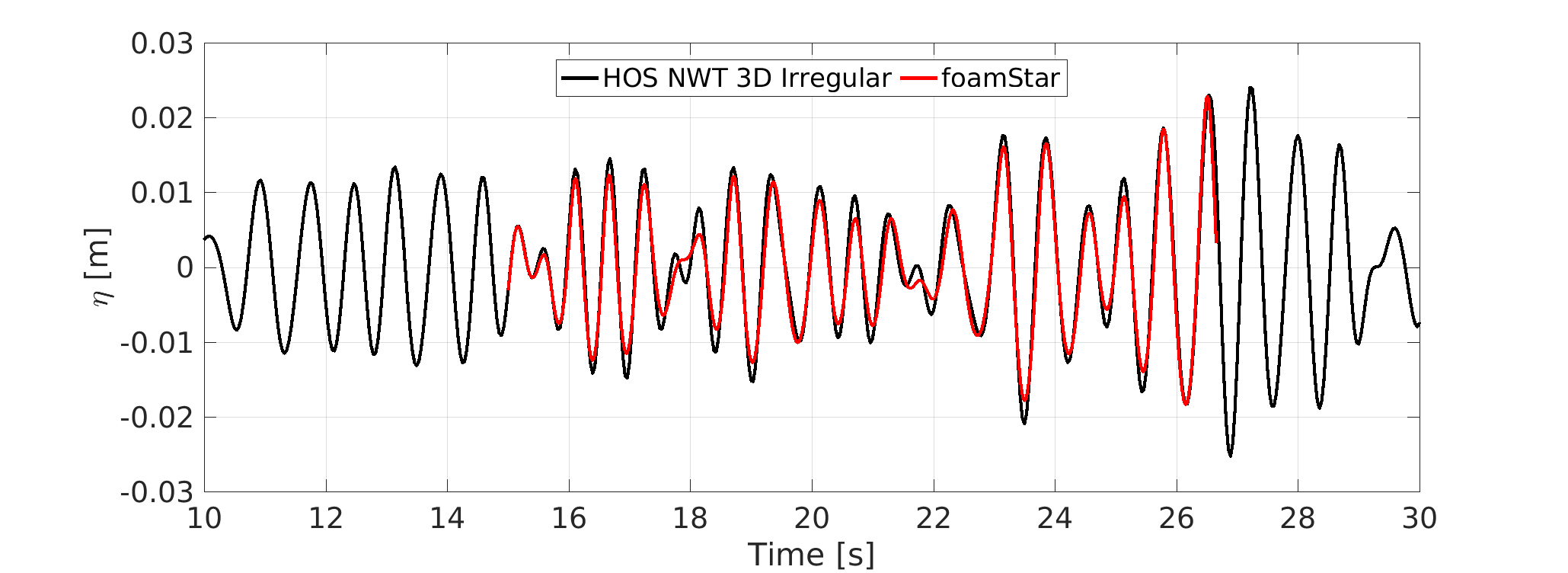}}
			\vspace{0.2cm}
			\caption{Comparison of HOS-NWT 3D wave elevation}
			\label{fig:HOSNWT_3DWaveElevation}
		\end{figure}
	}
	
	{
		\begin{figure} [H]
			\centering
			\subfloat[][ HOS-NWT 3D Regular Waves ] 
			{\includegraphics[scale=0.65]{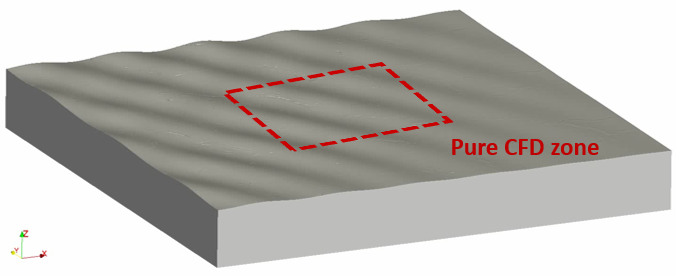}} \quad
			\subfloat[][ HOS-NWT 3D Irregular Waves]
			{\includegraphics[scale=0.65]{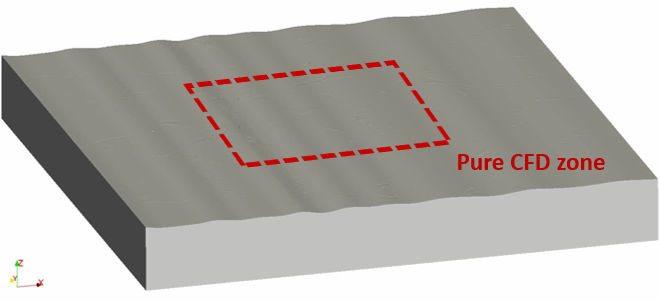}}
			\vspace{0.2cm}
			\caption{Snapshot of HOS-NWT 3D wave fields by \texttt{foamStar}}
			\label{fig:HOSNWT_3DSnapshot}
		\end{figure}
	}
	
	\pagebreak
	
	Simulation of extreme waves (1000 year return period waves in Gulf of Mexico(GOM) (Condition : $H_s=17.5 m, T_p=15.5 s, \gamma=3.3$) is simulated with HOS-NWT for 2D cases and used to generate waves in \texttt{foamStar}. The waves are compared with experiments performed in the wave basin of Ecole Centrale de Nantes (ECN). To simulate nonlinear breaking waves, the wave breaking model in HOS is utilized and allows also to capture when wave breaking occur. The expected wave breaking events are shown in Fig. \ref{fig:waveBreakingEvents}. In the experiments, wave breaking is observed at the expected position and time by HOS wave theory. The breaking moment in the experiment is shown in Fig. \ref{fig:waveBreakingExp}. The time series of wave elevation measured in the experiments are compared with the results of simulation using \texttt{Grid2Grid} in Fig \ref{fig:extremeWaveComparison}. The simulation snapshot when wave breaking occur is shown in Fig. \ref{fig:waveBreakingSimulation}. The small disturbance at the wave front is observed.
	
	{
		\begin{figure} [H]
			\centering
			\includegraphics[scale=0.7,trim=1.5cm 0 0 0.5cm]{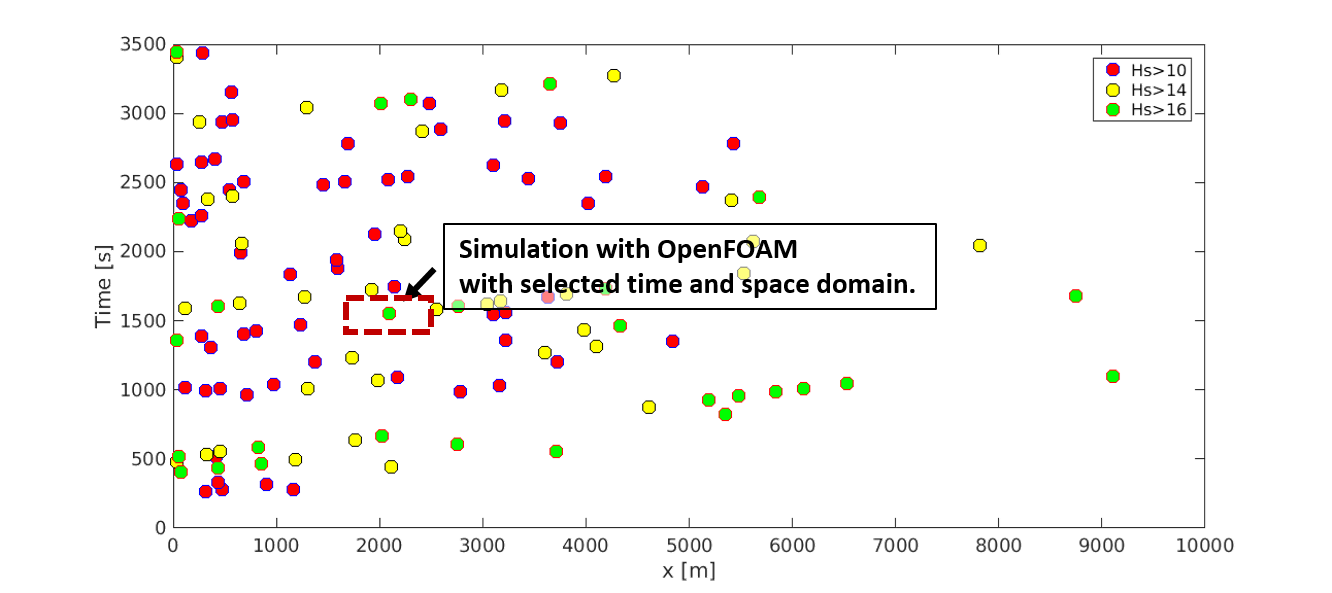}
			\caption{Expected wave breaking by HOS}
			\label{fig:waveBreakingEvents}
		\end{figure}
	}

	{
		\begin{figure} [H]
			\centering
			\includegraphics[scale=0.6,trim=1.5cm 0 0 1cm]{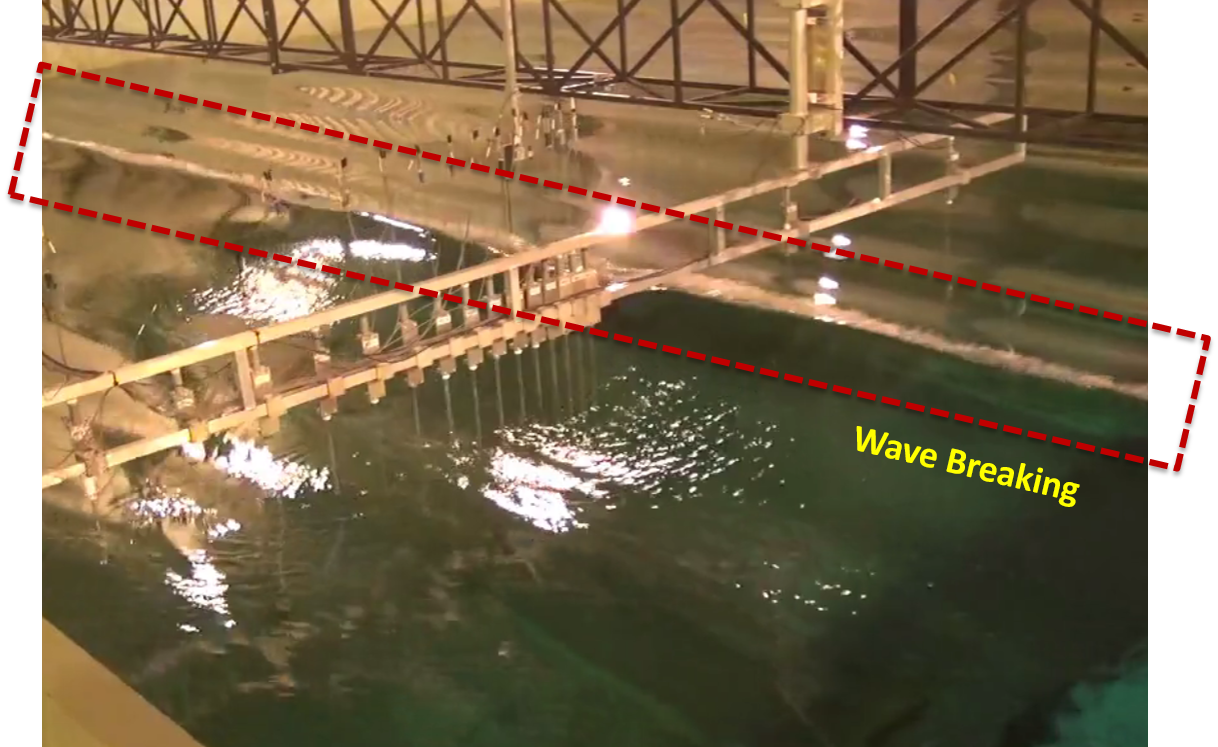}
			\vspace{0.1cm}
			\caption{Observed wave breaking in experiment}
			\label{fig:waveBreakingExp}
		\end{figure}
	}	
	
	\pagebreak	
	
	{
		\begin{figure} [H]
			\centering
			\subfloat[][ Overall wave time series ] 
			{\includegraphics[scale=0.21, trim ={1cm 0 0 5.3cm},clip]{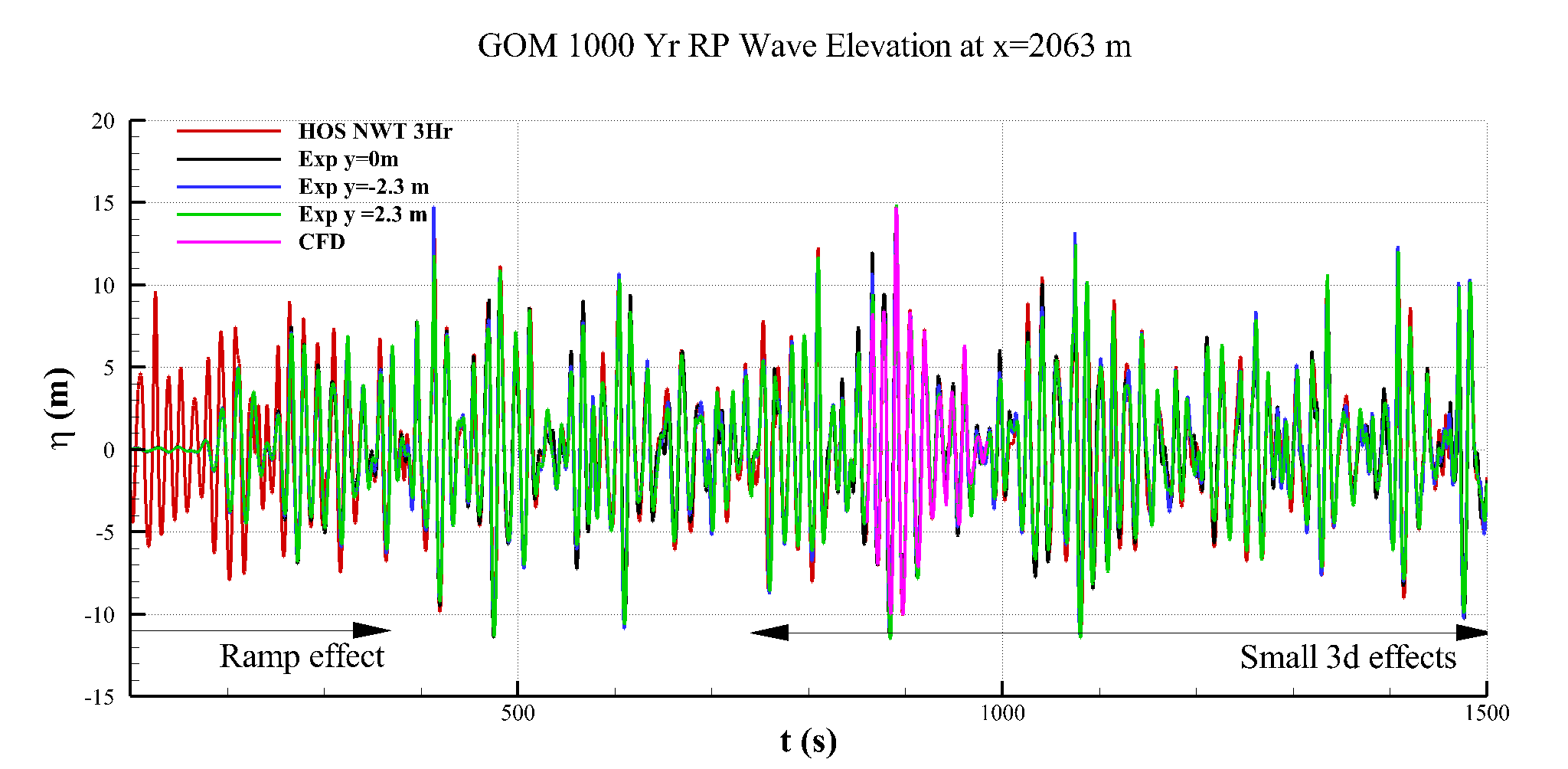}} \quad
			\subfloat[][ Wave elevation at the wave breaking moment]
			{\includegraphics[scale=0.21, trim ={1cm 0 0 3.32cm},clip]{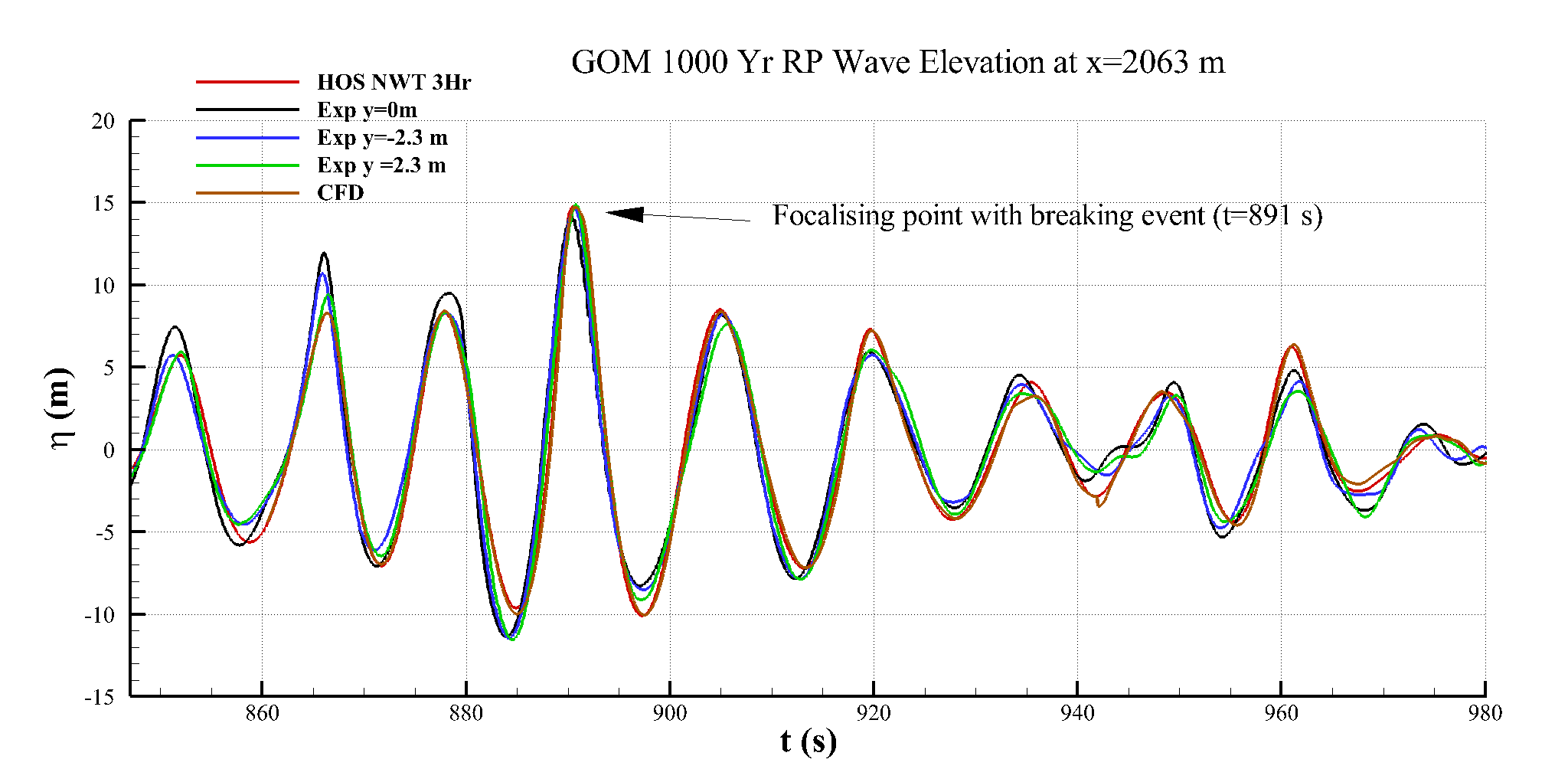}}
			\vspace{0.2cm}
			\caption{Snapshot of HOS-NWT 3D wave fields by \texttt{foamStar}}
			\label{fig:extremeWaveComparison}
		\end{figure}
	}
	
	{
		\begin{figure} [H]
			\centering
			\includegraphics[scale=0.6,trim=0 0 0 0]{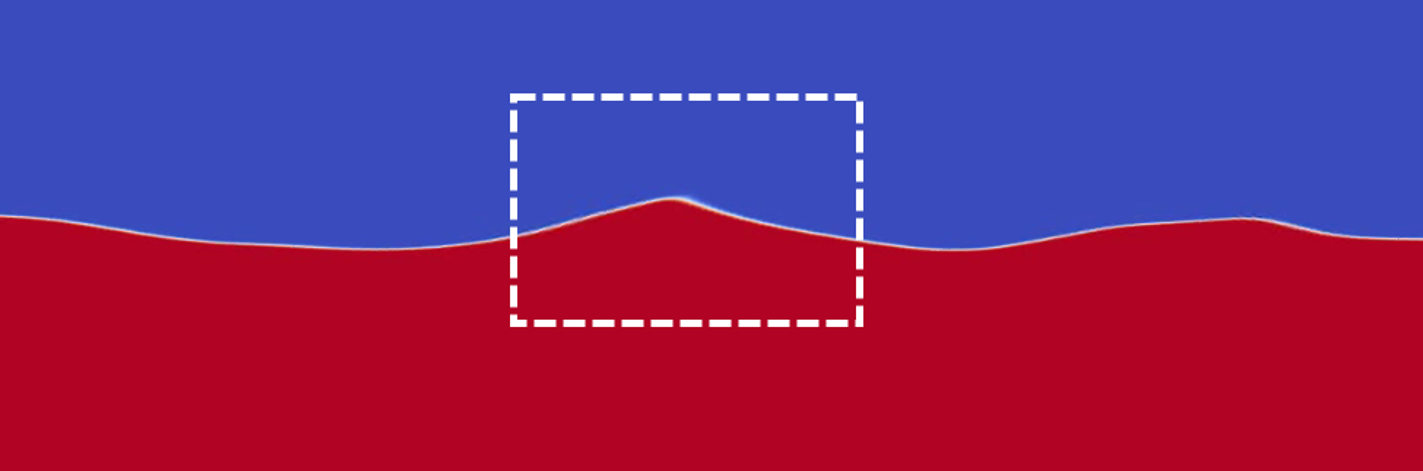}
			\vspace{0.3cm}
			\caption{Simulation of wave breaking by \texttt{foamStar} and \texttt{Grid2Grid}}		
			\label{fig:waveBreakingSimulation}
		\end{figure}
	}		
		
	\pagebreak
	\section{Summary}
	
	A HOS wrapper program called \texttt{Grid2Grid} is developed for the nonlinear wave simulation of numerical solvers. Most of data and functionality is encapsuled as a class to be easily used and maintained. \texttt{Grid2Grid} generates dynamic linked library as an independent package to be called and easily used in other languages. 
	
	The post processing of HOS is also possible by using \texttt{Grid2Grid}. Included post processing program is called \texttt{postGrid2Grid}. The usage is explained in Chapter \ref{chap:postGrid2Grid}.
	
	\texttt{Grid2Grid} is validated by using the code \texttt{foamStar} based on standard multiphase solver of OpenFOAM and also with an experiment. In the experiment, waves corresponding to 1000 year return period in Gulf of Mexico are generated. The wave elevation is measured at the wave breaking position expected by HOS wave theory and compared with the results of the simulation. Good agreement is shown between the measurement and the numerical solutions and the nonlinear wave phenomenon is observed both in experiment and in simulation. 
	
	In this document, the \texttt{Grid2Grid} program architecture, class and module structure, principle class data and functionality are explained to understand the feature of \texttt{Grid2Grid} and to be easily applied to numerical solvers. The interface examples with other programming languages are given as a source code and also in \texttt{Grid2Grid} package. 	
	

	\pagebreak
	\bibliography{reference}

	\bibliographystyle{apalike}

\end{document}